\documentclass[fleqn,usenatbib]{mnras}

\usepackage{newtxtext,newtxmath}
\usepackage[T1]{fontenc}

\DeclareRobustCommand{\VAN}[3]{#2}
\let\VANthebibliography\thebibliography
\def\thebibliography{\DeclareRobustCommand{\VAN}[3]{##3}\VANthebibliography}

\usepackage{graphicx}	% Including figure files
\usepackage{amsmath}	% Advanced maths commands
\usepackage{shortcuts}
\usepackage[capitalise]{cleveref}
\usepackage{ulem}
\usepackage{color,soul}
\Crefname{equation}{Eq.}{Eqs.}

\defcitealias{Kroupa2001}{K01}
\defcitealias{Jerabkova2018}{J18}
\defcitealias{MoeDistefano2017}{MD17}
\defcitealias{Chruslinska2019}{C19}
\defcitealias{Chruslinska2020}{C20}

\title[Binary Object environment-Sensitive Sampling]{Compact object populations over cosmic time I. \texttt{BOSSA}: a Binary Object environment-Sensitive Sampling Algorithm}

\author[L. M. de S\'a et al.]{
Lucas M. de S\'a,$^{1}$\thanks{E-mail: lucasmdesa@usp.br}
Ant\^onio Bernardo,$^{1}$
L\'ivia S. Rocha,$^{2}$
Riis R. A. Bachega,$^{1}$
Jorge E. Horvath,$^{1}$
\\
$^{1}$Instituto de Astronomia, Geof\'isica e Ci\^encias Atmosf\'ericas, Universidade de S\~ao Paulo, Rua do Mat\~ao, 1228, Butant\~a, S\~ao Paulo, BR \\
$^{2}$Instituto de F\'isica, Universidade of S\~ao Paulo, Rua do Mat\~ao, 1371, Butant\~a, S\~ao Paulo, BR
}

\date{Accepted XXX. Received YYY; in original form ZZZ}

\pubyear{2023}

\begin{document}
\label{firstpage}
\pagerange{\pageref{firstpage}--\pageref{lastpage}}
\maketitle

% Abstract of the paper
\begin{abstract}

Binary population synthesis (BPS) is an essential tool for extracting information about massive binary evolution from gravitational-wave (GW) detections of compact object mergers. It has been successfully used to constrain the most likely permutations of evolution models among hundreds of alternatives, while initial condition models, in contrast, have not yet received the same level of attention. Here, we introduce \texttt{BOSSA}, a detailed initial sampling code including a set of 192 initial condition permutations for BPS that capture both "invariant" and "varying" models, the latter accounting for a possible metallicity- and star formation rate (SFR)-dependence of the initial mass function (IMF); as well as correlations between the initial primary mass, orbital period, mass ratio and eccentricity of binaries. We include 24 metallicity-specific cosmic star formation history (cSFH) models and propose two alternate models for the mass-dependent binary fraction. We build a detailed pipeline for time-evolving BPS, such that each binary has well-defined initial conditions, and we are able to distinguish the contributions from populations of different ages. We discuss the meaning of the IMF for binaries and introduce a refined initial sampling procedure for component masses. We also discuss the treatment of higher-order multiple systems when normalizing a binary sample. In particular, we argue for how a consistent interpretation of the IMF implies that this is not the distribution from which any set of component masses should be independently drawn, and show how the individual IMF of primaries and companions is expected to deviate from the full IMF.
\end{abstract}

% Select between one and six entries from the list of approved keywords.
% Don't make up new ones.

%%%%%%%%%%%%%%%%%%%%%%%%%%%%%%%%%%%%%%%%%%%%%%%%%%

%%%%%%%%%%%%%%%%% BODY OF PAPER %%%%%%%%%%%%%%%%%%

\section{Introduction}
\label{sec:1intro}
Since the beginning of operations, and especially after upgrades were gradually implemented, the LIGO, Virgo and KAGRA gravitational-wave detectors have continuously extended the redshift threshold for observing compact object mergers. Released in 2019 by the LIGO-Virgo Collaboration, the first Gravitational Wave Transient Catalog \citep[GWTC-1,][]{GWTC1} included observations from both the first (O1) and second (O2) runs of the upgraded Advanced LIGO \citep{ALIGO}, the latter run O2 also joined by Advanced Virgo \citep{advanced_virgo}. At the time of the release of GWTC-1, the farthest observed merger was GW170729, at $z=0.49^{+0.19}_{-0.21}$. First released in 2020 as GWTC-2 \citep{GWTC2}, and then refined in 2021 as GWTC-2.1 \citep{GWTC21}, the second catalog contained new observations from Advanced LIGO and Advanced Virgo from the first half of the third run (O3), while those from the second half were released in 2022 as GWTC-3 \citep{GWTC3} by the LIGO-Virgo-KAGRA Collaboration (LVK). GWTC-2.1 and GWTC-3 included over $10$ new events at higher redshifts than GW170729, with GW190493 at $z=1.18^{+0.73}_{-0.53}$, and GW200220\_061928 at $z=0.90^{+0.55}_{-0.40}$, respectively. The KAGRA detector \citep{kagra2012,kagra2013} begun observations during a two-week extension of the third run, O3GK \citep{O3GK}, alongside GEO600 \citep{GEO600_2010,GEO600_2016}, and will continue to join future runs.

We are now in the middle of the second half of LVK's fourth observing run (O4b), which started in April 2024 and is expected to run until June 2025. Throughout the first (O4a) and second halves, both the Hanford and Livingston LIGO detectors have been active \citep[now with the A+ plus upgrade,][]{APLUS}, with the upgraded Advanced Virgo \citep[AdV+,][]{VirgoPlus} joining O4b, and KAGRA, which operated for the first month of O4a, expected to join observations by early 2025 \citep[including the KAGRA+ upgrade series,][]{KAGRAplus}. Further upgrades are already planned for the near future and should integrate the O5 run, currently  projected to begin in 2027. One such upgrade is LIGO Voyager \citep{Voyager}, with an expected threshold at $z\sim 8$ for black hole-black hole (BHBH) merger detections, while the long term plans contemplate the construction of the LIGO-India detector \citep{LIGO-India}; third-generation observatories such as Cosmic Explorer \citep{CosmicExplorer,CosmicExplorer2023} and the Einstein Telescope \citep{EinsteinTelescope}, which will further raise the threshold up to redshift $\sim 10$ for black hole-neutron star (BHNS) mergers, and even beyond for BHBHs; as well as the space-based Laser Interferometer Space Antenna \citep[LISA,][]{LISA2017,LISA2023,BREIVIKLISA}, with a frequency range projected to extend to inspiraling compact binaries (including those hosting white dwarfs), extreme mass ratio inspirals and massive black hole binaries. When accomplished, the upgrade and extension of the roster of gravitational-wave (GW) observatories will bring the overall range of GW detections to around the age of the first stars, and encompass all compact object mergers ever produced. As we approach this level of coverage, any possible variations of the population properties, such as their merger rates and mass distributions, with redshift should become apparent, and modeling of these variations will come to constitute a fundamental ingredient for both predicting future observations and extracting information about binary formation and evolution from actual observations.

Even at current sensitivities, a possible redshift-dependence of compact binary merger (CBM) populations may have significant effects, due to the coalescence time of binary compact objects (BCOs), which can range anywhere from $\sim1\,\mathrm{Myr}$ to over a Hubble time. This implies that even some of the nearest merging binaries could have formed and evolved under very different circumstances than those observed locally. The main quantity expected to considerably affect the properties of a given population is the metallicity of the environment in which it formed: in terms of evolution, metallicity is a defining factor for the wind mass loss efficiency \citep[e.g.,][]{Hurley2000,vink200wind,Vink2006,belczynski2010winds}, affects the radius of massive stars and thus the interaction between binary companions \citep[e.g.][]{Hurley2000,KlenckiRadii2020}, influences the outcome of supernova explosions \cite[e.g.,][]{FryerSN2012,fryer2022supernova,patton2020supernova,schneider2021supernova,burrows2024supernova}, and affects the balance between stable and unstable mass transfer \citep[e.g.,][]{marchant,gallegos}. While in terms of initial conditions, there is a longstanding expectation that a relative excess of massive stars should form in low-metallicity environments, relative to those of Solar metallicity \citep[see][for extensive reviews of the subject]{Kroupa2013,Hopkins2018}. Metallicity should also affect pre-zero age main sequence (ZAMS) evolution, and thus the orbital parameters of ZAMS binaries, to some extent \citep[e.g.,][]{El-BadryBinFrac2018,MoeBinFrac2019}. 

An essential tool in exploring the implications of such binary formation/evolution uncertainties for CBM populations is binary population synthesis (BPS), which allows extensive testing of the available models in a practical time through the implementation of fits and analytical prescriptions. These methods are today implemented by a variety of codes, under different approaches, such as \texttt{binary\_c} \citep{izzard2004,izzard2006,izzard2009,izzard2018}, \texttt{BPASS} \citep{eldridge2017,stanway2018,byrne2022}, the Brussels code \citep{brussels1998,brussels2004,brussels2014}, \texttt{MCLUSTER} \citep{kamlah2022}, \texttt{COMBINE} \citep{Kruckow2018}, \texttt{COMPAS} \citep{stevenson2017,vignagomez2018,teamCOMPAS2022}, \texttt{COSMIC} \citep{Breivik2020}, \texttt{MOBSE} \citep{Mapelli2018,Giacobbo2018}, \texttt{POSYDON} \citep{POSYDON}, the Scenario Machine \citep{Lipunov1996,Lipunov2009}, \texttt{SEBA} \citep{SEBA1996,SEBA2012}, \texttt{SEVN} \citep{SEVN2019,SEVN2020} and \texttt{StarTrack} \citep{STARTRACK1,STARTRACK2,STARTRACK3}.

The metallicity-dependent cosmic star formation history (cSFH) is also an important quantity when studying populations that present a wide range of ages, as it describes how much mass was converted into stars, and at what rate, at different moments throughout the history of the Universe \citep[see e.g.,][]{SFHMadau,artale2019,Chruslinska2019,Chruslinska2020}. All of these factors must, in principle, be taken into account in BPS in order to accurately reproduce the properties of the physical population of CBMs, or more generally BCOs, and for a given set of initial conditions and evolutionary models, predict the redshift-dependent merger rate and other population properties \citep[see, e.g.,][]{Neijssel2019,santoliquido2021,Boco2021,broekgaarden2022,vanSon2023,chruslinska2024}. Once selection effects for observable mergers have been taken into account, the synthetic local properties can be compared to observations for validation, constraining the most likely combinations of initial conditions end evolutionary models; and from a validated permutation of models, predictions for future upgraded or enhanced new facilities can be made.

The metallicity-dependent cSFH, however, should not be the only source of time-dependence for BPS. Besides the already mentioned metallicity-dependent evolution models (which tie into the increasing mean metallicity of galaxies with time), the process of star formation is also expected to vary, both through a metallicity-dependence and otherwise. While \citet{deMink2015} and \citet{Kruckow2018} had already investigated the effect of IMF variations on CBM populations, \citet{Klencki2018} pointed toward the further step of treating the initial conditions themselves as environment-dependent. There is a longstanding expectation that the IMF should produce an excess of massive stars at low metallicities in relation to the original \citet{Salpeter1955} IMF, and its "descendants" \citep{SalpeterDescendantsKroupa}, which has traditionally been argued to stem from the dependence of the Jeans mass on temperature, and the increasing efficiency of cooling with metallicity \citep{Larson1998,Larson2005,Bate2005,Bonnell2006}; or from self-regulating pre-ZAMS accretion \citep{adams1996selfacc,Matzner2000selfacc,federrath2014selfacc,federrath2015selfacc}. In spite of the argument having stood for decades, observations have continuously challenged it, and reinforced the notion that the IMF is essentially invariant \citep[see, e.g., the review by][]{Bastian2010IMF}. On the other hand, it has been argued that systematic issues in the study of the IMF led to an overestimation of the agreement between different measurements of the IMF, which are in fact not enough to ascertain that it is invariant \citep{Hopkins2018}. Over more than fifty years after the original work by \citet{Salpeter1955}, the previous decade has seen the emergence of new evidence toward IMF variations \citep[e.g.,][]{Schneider2018Doradus}, and in particular toward a systematically varying IMF not only with metallicity \citep{MarksIMF2012}, but also with the SFR \citep{Gunawardhana2011}. This observational evidence prompted the construction of new models which reproduce the local \citeauthor{Salpeter1955} IMF, but show variations in different environments, especially within the results of the integrated galaxy-wide IMF framework \citep{KroupaWeidner2003,Yan2017,Jerabkova2018}, which also ties to ongoing debate around the nature of star formation as a stochastic or strongly self-regulating process \citep[e.g.][]{Eldridge2012stoch,Weidner2013_mmax,Weidner2014,Stanway2023stoch} and other current uncertainties around star formation \citep[see the reviews by][]{Kroupa2013,Kroupa_Jerabkova_2021}.

Although some support has been found, variations of other relevant initial parameters --- orbital period, mass ratio and eccentricity --- are yet too uncertain in terms of observational evidence, in order to construct environment-dependent distributions for them. Until recently, some of the most common choices for the sampling of these parameters remained \"Opik's law for orbital period/semi-major axis \citep{OpikLaw,abtOpiklaw} and the \citet{SanaMassRatio} mass ratio distribution. Although a log-uniform distribution for ZAMS orbital periods remained commonplace, \citet{SanaMassRatio} had already determined a strong preference for close pairs among massive O-type stars. The constraining of correlations between mass and orbital parameters at ZAMS --- including a preference for close pairs for increasingly massive stars ---, was again taken up by \citet{MoeDistefano2017}, who found that these distributions are best fitted by power-law and linear function series, and that they are strongly correlated. In particular, all three are correlated with the mass of the primary component of the binary. If the primary mass is distributed according to an environment-dependent IMF, this could in principle provide a first glance at systematic variations of the orbital parameters with environmental conditions, although this might not necessarily be significant compared to other sources of uncertainty. The effects of the correlated orbital parameters from \citet{MoeDistefano2017} in BCO populations were studied in the aforementioned work by \citeauthor{Klencki2018}, employing the metallicity-dependent IMF fit by \citet{MarksIMF2012}, who found that these models only affect the merger rates resulting from BPS by a numerical factor of $2.1-2.6$.

The role of higher-order multiple systems is also significant for generating CBMs, and a series of works have been dedicated to their particular contribution to the population of CBMs \citep[e.g.,][]{Fragione2020hierarch_spin,Vynatheya2022hierach,stegmann2022triples}. To some extent, BPS can account for contributions from inner binaries if we can set apart those higher-order multiples and assume the inner binaries to evolve essentially independently, a topic also addressed by \citet{Klencki2018}. This approach can help establish a lower limit for merger rates. Alternatively, as often assumed, all multiples can be treated as binaries to estimate CBMs originating from multiples of any order.

In an effort to account for all those concerns, in this paper we present \texttt{BOSSA} (Binary Object environment-Sensitive Sampling Algorithm)\footnote{\url{https://github.com/lmdesa/BOSSA}}, an initial sampling algorithm that provides an expressive refinement to the entire initial sampling process as it is performed for BPS. While our ultimate interest is to provide new estimates for the time-evolution of CBM properties, here we present the construction of \texttt{BOSSA} as a general tool for BPS, and reserve the full discussion of our main results for mergers to a second work. We account for both the metallicity- and SFR-dependent IMF from \citet{Jerabkova2018} and the correlated orbital parameters from \citet{MoeDistefano2017}, with a careful treatment of the higher-order multiples present in the latter's sample. These distributions were coupled to IGIMF-corrected, when necessary, SFR and metallicity distributions over redshift, in order to make this sampling "time-sensitive" from the start. We are careful with our treatment of the assumed probability distributions, so that their physical interpretation in each case is clear, and the treatment accordingly transparent. We also include technical details about the minimizing procedure of the number of populations that need to be run through a BPS code in order to obtain a good sample of the evolved population.

We present the \texttt{BOSSA} pipeline as follows. In Sec. \ref{sec:2distributions}, we introduce and briefly describe the initial conditions distributions that have been collected, and discuss their physical interpretation and means of implementation where appropriate. Then, in Sec. \ref{sec:3sampling} we describe our pipeline employing these distributions; we also go into detail on variations of the initial sampling with regard to component masses, and describe our suggested methods for dealing with these variations. We explore the resulting sample in Sec. \ref{sec:4results},  checking the consistency of our mass sampling and looking for emerging correlations between orbital parameters, as well as multiplicity fractions, and redshift or metallicity. Finally, in Sec. \ref{sec:5conclusions} we summarize our work and offer our concluding remarks. In an accompanying paper \citep{bossa2}, we apply our pipeline to the synthesis of a population of CBMs and compare the results for two representative initial conditions model permutations.
\section{Distributions}
\label{sec:2distributions}

In the most general case, the initial sampling involves eight quantities. First, the mass of the primary component ($m_1$) and its number of companions ($\ncp$). Second, for each companion, a set of three orbital parameters: the orbital period ($P$), the mass ratio ($q=m_\mathrm{cp}/m_1\leq1$, $m_\mathrm{cp}$ the companion mass), and the eccentricity ($e$). And, finally, the environmental parameters: metallicity ($\Z$), star formation rate ($\SFR$) and redshift ($z$). We broadly divide the distributions from which we sample these quantities into the IMF itself, discussed in Sec. \ref{sec2sub:imf}; the orbital parameter distributions, discussed in Sec. \ref{sec2sub:orbital_parameters}; and the environmental distributions, discussed in Sec. \ref{sec2sub:environment}.

\subsection{The initial mass function}
\label{sec2sub:imf}

In simple terms, the IMF can be defined as the probability density function of masses of a group of stars born in the same stellar formation event. It was first measured by \citet{Salpeter1955} as a power-law in the range $0.4\lesssim m/\Msun \lesssim 10$ with slope $\alpha=-2.3$, which has become known as the Salpeter index, or Salpeter-Massey index, in reference to \citet{Massey1998} who later showed that the Salpeter index extends to $m\gtrsim10\,\Msun$. Further work found that the IMF displays a flatter slope at lower masses, typically $m\lesssim0.5\,\Msun$ \citep[e.g.][]{MillerScaloIMF,ChabrierIMF}, with, namely, the IMF by \citet{Kroupa2001} (K01 hereon) becoming widely used in BPS. More recently, the inclusion of the brown dwarf range in the IMF has given it a positive slope below the hydrogen burning limit of $0.08\,\Msun$ \citep{Kroupa2013}.

For as long as the IMF has been studied, one of the most contentious questions surrounding it has been of whether it is invariant, i.e., if the IMF is the same for every star formation event, independently of the environment. The most traditional argument in favor of a \textit{varying} IMF has been based on the role presumably played by Jeans' mass in the stellar formation process,

\begin{equation}
    M_\mathrm{J} \propto \rho^{-1/2}T^{3/2},
\end{equation}

\noindent where $\rho$ is the density and $T$ the temperature of the gas; $M_\mathrm{J}$, being proportional to a power of the temperature, should increase with decreasing metallicity, since cooling is less efficient at low metallicities \citep{Larson1998,Larson2005,Bate2005,Bonnell2006}. An alternative argument, based on self-regulating pre-ZAMS accretion, results in the same trend between newborn star masses and metallicity \citep{adams1996selfacc,Matzner2000selfacc,federrath2014selfacc,federrath2015selfacc}. On the other hand, the IMF has been observed to be consistent with the \citet{Salpeter1955} IMF in many different environments, while evidence for variation has been more sparse. Alternatively, a universal IMF may be explained as a consequence of the interplay between different mechanisms at play in stellar formation such as turbulence, magnetic fields, stellar feedback and the composition of the original environment \citep[dust, in addition to gas; see the reviews by][]{imfreview2014offner,imfreview2024hennebelle}. Of these, turbulence is frequently pointed to as a universal mechanism for setting the IMF, in connection to the core mass function (CMF) that is expected to yield the IMF after collapse \citep[see][for the connection between the CMF and IMF, and the role of turbulence in setting the Salpeter slope]{imfturbulence2015guszejnov}. Star formation simulations have also shown the efficiency of self-regulating mechanisms in keeping it invariant over metallicity \citep[e.g.][]{IMFBate2014,IMFBate2019}, down to $10^{-5}\Zsun$ \citep{Peters2014}. The relative lack of compelling evidence for variations and the traditional preference for the simplest possible solution have thus led to a well-established assumption of a "universal" IMF \citep{Hopkins2018}, but it continues to be challenged. Within the last two decades, in particular, enough evidence for variations has been accumulated to support empirically-based models for a varying IMF \citep[see][for a summary]{Kroupa2020}.

We will be concerned here with a possible dependence of the IMF on two quantities: the metallicity ($\Z$) and the SFR. For the first there has been evidence based on Galactic globular cluster (GC) observations: \citet{DeMarchi2007} first noticed for a sample of 20 Galactic GCs a trend of the \textit{present-day} mass functions (PDMF) becoming steeper at lower masses for higher-concentration GCs, something which \citet{Marks2008} argued could be explained by a scenario in which the missing low-mass stars had been lost by the GCs due to feedback from massive stars concentrated in the core of a mass-segregated cluster. \citet{MarksIMF2012} followed this up by developing a model which explicitly connects the loss of low-mass stars to a top-heavy IMF, yielding a model for an IMF which becomes top-heavy (shows an excess of massive stars relative to \citet{Salpeter1955}) at low metallicities.

Regarding the SFR, \citet{Gunawardhana2011} analyzed a sample of $\sim40,000$ galaxies from the Galaxy and Mass Assembly (GAMA) survey and found a strong correlation between the IMF power law index at high masses and the SFR, which could be directly fitted to a high-mass index increasingly flatter with increasing SFR, making the IMF top-heavy at high SFR.

We work with two models for the IMF. In the "Invariant IMF" model, we use the \citetalias{Kroupa2001} IMF, in line with most BPS works. In the "Varying IMF" model, we use the metallicity- and SFR-dependent IMF developed by \citet{Yan2017} and \citet{Jerabkova2018} within the IGIMF theory, based on the empirical fits by \citet{Gunawardhana2011} and \citet{MarksIMF2012}. In Sec. \ref{sec2sub:imf_conventions}, we establish some conventions and good practices regarding the IMF; in Sec. \ref{sec2sub:igimf} we discuss the IGIMF model in \citetalias{Jerabkova2018}.

\subsubsection{Conventions}
\label{sec2sub:imf_conventions}

Once we admit that the IMF may vary, it becomes important to introduce a terminology that specifies the environment to which a particular IMF is thought to apply to. We will refer to the IMF within a singular star-forming region as the \textit{stellar} IMF, or sIMF ($\xi_\mathrm{s}$), while the "effective" stellar IMF of a galaxy we refer to as the \textit{galaxy-wide} IMF, or gwIMF ($\xi_\mathrm{gw}$). 

In keeping with the recommendations by \citet{Hopkins2018} to avoid confusions in conventions and definitions that have often afflicted the study of the IMF, we define the \textit{slope}, $\alpha$, to always carry the sign in a power law or power law-series IMF,

\begin{equation}
    \label{sec2eq:powerlaw_imf_def}
    \xi\lrp{m} \propto m^\alpha.
\end{equation}

We keep our convention even when it is defined otherwise in referenced works. For the stellar IMFs, unless stated otherwise, we exclude brown dwarfs and set $m_\mathrm{min}=0.08\,\Msun$, and $m_\mathrm{max}=150\,\Msun$ for the upper stellar mass limit (e.g., \citeauthor{figer2005max_ast_mass}, \citeyear{figer2005max_ast_mass}; \citeauthor{Oey2005max_ast_mass}, \citeyear{Oey2005max_ast_mass}; \citeauthor{Koen2006max_ast_mass}, \citeyear{Koen2006max_ast_mass}; \citeauthor{Maiz2007max_ast_mass}, \citeyear{Maiz2007max_ast_mass}; cf. \citeauthor{Schneider2018Doradus}, \citeyear{Schneider2018Doradus}; \citeauthor{Shenar2023max_ast_mass}, \citeyear{Shenar2023max_ast_mass}). Although the labels "low-mass" and "high-mass" are often deliberately imprecise, we define them to mean the $<1.0\,\Msun$ and $\geq1.0\,\Msun$ regions, respectively, when referring to the stellar mass function slopes.

When speaking about variations of the IMF, they are described in reference to the \citet{Salpeter1955} IMF: an IMF with a relative excess (deficiency) of high-mass stars is said to be {\it top-heavy} (light); while one with a relative excess (deficiency) of low-mass stars is said to be {\it bottom-heavy} (light). Note that top-heaviness does not imply bottom-lightness (or vice-versa).

\subsubsection{Integrated galaxy-wide IMF (IGIMF) theory}
\label{sec2sub:igimf}

\begin{figure*}
    \centering
    \includegraphics[width=\textwidth]{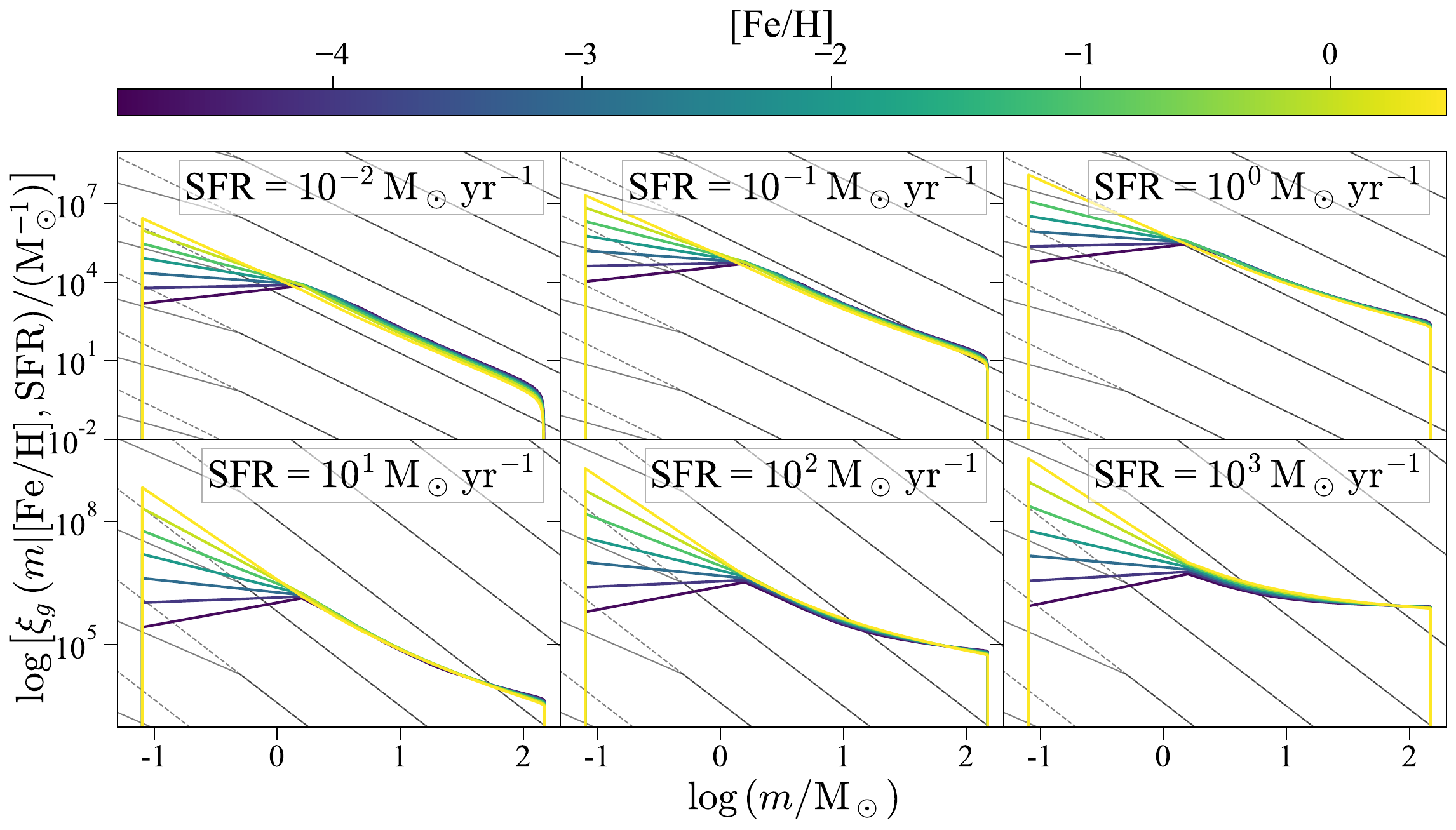}
    \caption{The gwIMF (equation \eqref{eq:gIMF_final}) computed following \citetalias{Jerabkova2018} (thick colored lines) for different SFR and metallicity combinations, here treated as independent variables. The \citet{Salpeter1955} (thin dashed black lines) and \citetalias{Kroupa2001} (thin solid black lines) are shown for comparison. The fits guarantee that the gwIMF becomes top-heavy and strongly bottom-light with decreasing metallicity; and strongly top-heavy with increasing SFR. Each curve is normalized to $\SFR\times10\,\mathrm{Myr}$.}
    \label{fig:sfr_imfs}
\end{figure*}

The IGIMF theory, originally due to \citet{KroupaWeidner2003}, connects the sIMF and gwIMF in a straightforward and intuitive manner: the gwIMF of a given galaxy is nothing more than the sum of the sIMFs of all of its star-forming regions, termed {\it embedded clusters} (ECLs). This approach was subsequently refined, and we present it here as it was employed by \citet{Yan2017}, in studying sampling methods (discussed in Sec. \ref{sec3sub:optimal_sampling}); and by \citet{Jerabkova2018} (\citetalias{Jerabkova2018} hereon), who built a model based on the fits by \citet{Gunawardhana2011} and \citet{MarksIMF2012}.

Operationally, the gwIMF results from the spatial integration of the sIMF over all ECLs in a galaxy, formed within a time interval $\delta t$. With regards to star formation, attributes such as mass and chemical composition of an ECL are relevant to setting the sIMF within, and expected to vary spatially over a galaxy. While it is possible to take into account this variation of composition within IGIMF theory \citep[see][]{PflammAltenburg2008}, we follow \citet{Yan2017} and \citetalias{Jerabkova2018} by assuming, as a first approximation, each galaxy to be chemically homogeneous. This is enough for our objective of capturing the variation of \textit{average} star-formation conditions over cosmic time.

As the most directly relevant parameter to stellar evolution and an explicit input condition to population synthesis codes, we take the metallicity $\FeHinline$ as a stand-in for galactic composition. For a galaxy of fixed metallicity, ECLs are differentiated solely by their total \textit{stellar mass}, $M_\ecl$, and the computation of the gwIMF becomes an integration over $M_\ecl$. Hence, if we define an \textit{ECL initial mass function}, or eIMF ($\xi_\mathrm{e}$), the gwIMF can be written as

\begin{align}
    &\xi_\mathrm{gw}\lrp{m\middle|\FeHinline,\mathbf{x}_\mathrm{g}} = \nonumber\\
    &\int_\mathrm{galaxy}\; \xi_\mathrm{s}\lrp{m\middle|M_\ecl,\mathbf{x}_\mathrm{{s}}}\; \xi_\mathrm{e}\lrp{M_\ecl\middle|\mathbf{x}_\mathrm{e}}\; \d M_\ecl, \label{eq:gimf_def}
\end{align}

\noindent in a general case, where $\mathbf{x}_\mathrm{s}$ and $\mathbf{x}_\mathrm{e}$ represents any galactic parameters on which the IMFs are found to depend, at least one of them being the metallicity. 

The IGIMF theory places two further constraints on the IMF (both sIMF and eIMF). The first is simply that of normalization: for a given total star mass $M_\mathrm{tot}$, it must be true that

\begin{equation}
    \label{eq:imfconstraint1_totmass}
    \int_{\mmin}^{\mmax}\; m\; \xi\lrp{m\middle|\mathbf{x}}\; \d m = M_\mathrm{tot},
\end{equation}

\noindent in the relevant physical $\lrs{m_{\min},m_{\max}}$ range, where, for the sIMF in a given ECL, $M_\mathrm{tot}=M_\ecl$, and for the eIMF, $M_\mathrm{tot}$ will depend on the chosen time interval $\delta t$ and galaxy SFH.

The second constraint is non-trivial and justified chiefly by the success it has provided to the IGIMF theory to reproduce observations \citep{Yan2017}. In its original formulation, it expresses the notion that there should be a single object between $\mmax$ and some empirical maximum mass $\mmax^\mathrm{emp}$, or

\begin{equation}
    \label{eq:imfconstraint2_mmax}
    \int_{\mmax}^{\mmax^\mathrm{emp}}\; \xi\lrp{m\middle|\mathbf{x}}\; \d m = 1.
\end{equation}

These two constraints together fully determined not only the IMF normalization but also $\mmax$, which is allowed to vary below $\mmax^\mathrm{emp}$. For the sIMF, we set $\mmax^\mathrm{emp}=150\,\Msun$, while for the eIMF we keep $M_{\mathrm{ecl,max}}^\mathrm{emp}=10^9\,\Msun$ from \citet{Yan2017}. This leads to the mass of the most massive star in a given ECL to be correlated to $M_\ecl$ \citep[the $m_\mathrm{max}-M_\ecl$ relation, see][]{Weidner2013_mmax}; and the mass of the most massive ECL in a galaxy to be correlated to its SFR \citep[the $M_{\ecl,\max}-\SFR$ relation, see][]{Weidner2004sfr}. This last case best fits the observed relation by assuming that, for the eIMF,

\begin{equation}
    \label{eq:mtot_sfr}
    M_\mathrm{tot} = \SFR\cdot\delta t,
\end{equation}

\noindent where we assume an episode of constant SFR of duration $\delta t=10\,\mathrm{Myr}$, which is consistent with independently constrained survival timescales of giant molecular clouds \citep{MeidtMC2015,PadoanMC2016}, and so far has led to the best agreement between IGIMF theory and the empirical $M_{\ecl,\max}-\SFR$ relation \citep[see Axiom 5 and Figure 2 of ][]{Yan2017}.

In order to bring the gwIMF from IGIMF theory in accordance with the SFR-dependence of the high-mass slope found by \citet{Gunawardhana2011}, \citet{Weidner2013_sfr} determined that this dependence must be inserted into the eIMF through either its minimum mass or slope. We follow their suggestion for the second case, as in \citetalias{Jerabkova2018}, in which the eIMF is defined as 

\begin{equation}
    \label{eq:eIMF}
    \xi_\mathrm{e}\lrp{M_\ecl\middle|\SFR} = \begin{cases}
         k_\ecl M^{\beta\lrp{\SFR}}, & 5\,\Msun\leq M_\ecl < M_\text{ecl,max}\lrp{\SFR}, \\
         0, & \text{otherwise},
    \end{cases}
\end{equation}

\noindent where $M_{\ecl,\max}$ is defined by equations \eqref{eq:imfconstraint1_totmass} and \eqref{eq:imfconstraint2_mmax}; and the index, in fitting to \citet{Gunawardhana2011}, is

\begin{equation}
    \label{eq:eIMF_beta}
    \beta\lrp{\SFR} = 0.106\log\frac{\SFR}{\Msun\text{ yr}^{-1}} - 2.
\end{equation}

The stellar IMF within an ECL, in the spirit of the \citetalias{Kroupa2001} IMF, is described by a power law series,

\begin{equation}
    \label{eq:sIMF}
    \xi_\mathrm{s}\lrp{m\middle|\FeHinline,M_\ecl} = \begin{cases}
         k_1m^{\alpha_1}, & 0.8\,\Msun \leq m < 0.50\,\Msun,  \\
         k_2m^{\alpha_2}, & 0.50\,\Msun \leq m < 1.00\,\Msun, \\
         k_3m^{\alpha_3}, & 1.00\,\Msun \leq m < \mmax\lrp{M_\ecl}, \\
         0, & \text{otherwise}..
    \end{cases}
\end{equation}

\noindent where all indices are explicitly dependent on $\FeHinline$ and the normalization constants implicitly dependent on both $\FeHinline$ and $M_\ecl$ through equations \eqref{eq:imfconstraint1_totmass} and \eqref{eq:imfconstraint2_mmax}.

The high-mass index, $\alpha_3$, is set to the empirical fit from \citet{MarksIMF2012}, which \citetalias{Jerabkova2018} showed to be more conveniently written as 

\begin{equation}
\label{ch1eq:marks_alpha3}
    \alpha_3 = \left\{\begin{array}{lr}
         2.3, & \text{if }x<-0.87,  \\
         -0.41x+1.94, & \text{if }x\geq-0.87, 
    \end{array}\right.,
\end{equation}

\noindent where

\begin{equation}
    \label{eq:second_x_relation}
    x=-0.14\FeH + 0.6\log\lrp{\frac{M_\text{ecl}}{10^6M_\odot}}+2.83.
\end{equation}

For the other two exponents of the stellar IMF, \citet{MarksIMF2012} suggest the following empirical relation,

\begin{equation}
    \label{eq:kroupa_a23}
    \alpha_i = \alpha_{ic} + \Delta\alpha\FeH,
\end{equation}

\noindent where $\Delta\alpha\approx0.5$ and $\alpha_\text{ic}$ are the respective exponents of the Invariant IMF. The relation is based on a rough estimate by \citet{Kroupa2002} for stellar populations in the Milky Way disc, the bulge, and globular clusters spanning a range of about $\left[\text{Fe}/\text{H}\right]=0.2$ to $\approx-2$, with values beyond this range based on extrapolation. This relation leads to a bottom-light IMF at low metallicities.

Finally, the gwIMF is computed as 

\begin{align}
    \xi_\mathrm{gw}\lrp{m\middle|\FeHinline,\SFR}& = \nonumber\\
    \int_5^{10^9} &\xi_\mathrm{s}\lrp{m\middle|\FeHinline,\SFR}\;\xi_\mathrm{e}\lrp{M_\ecl\middle|\SFR}\;\frac{\d M}{\Msun}, \label{eq:gIMF_final}
\end{align}

\noindent which results in the curves shown in Fig. \ref{fig:sfr_imfs} as examples, reproducing both the fitted behaviors with SFR and with metallicity.

%%%%%%%%%%%%%%%%%%%%%%%%%%%%%%%%%%%%%%%%%%%%%%%%%%%%%%%%%%%%%%%%%%%%%%%%%%%%%%%%%%%

\subsection{Orbital parameter distributions}
\label{sec2sub:orbital_parameters}

The remaining binary parameters -- orbital period ($P$), mass ratio ($q=\mcp/m_1\geq1$) and eccentricity ($e$) --, which we collectively refer to as \textit{orbital parameters}, have been often modeled by \"Opik's law \citep{OpikLaw}, a log-uniform distribution between $10^{0.2}$ and $10^{3}\mathrm{d}$ for $P$; a uniform distribution for $q$ \citep{SanaMassRatio}, usually in the range $[0.01,1]$; and assumed circular orbits, i.e., $e=0$. We adopt this set of initial conditions as our \textit{Invariant} (uncorrelated) orbital parameters due to their commonality, but do highlight that other distributions have been in use in the community for the past decade, such as the $\propto\logP^{0.55}$ and $\propto e^{0.41}$ distributions already constrained by \citet{SanaMassRatio}.

For what we will refer to as \textit{Varying} orbital parameters, we employ the distributions by \citet{MoeDistefano2017} (\citetalias{MoeDistefano2017} hereon), who studied data from over 20 surveys of massive multiple systems, each covering a narrow range of orbital periods and masses, carefully corrected for incompleteness due to observational biases and then fitted with analytic distributions (power law and linear expressions) to the full set of data. These distributions revealed considerable correlations between the orbital parameters, which, as the authors carefully discuss, can be explained in terms of the pre-ZAMS evolution of multiple systems. Here we will simply relate the form of the distributions considered and discuss their implementation where we deviate from the original results; we only briefly reproduce some of the proposals for the physical origin of the shape of the distributions, and refer to the original publication (and references therein) for a full discussion.

The full set of data covers the $0.1<q<1.0$, $0.2<\log \lrp{P/1\text{ d}}<8.0$ and $0.8\leq m_1/\Msun<40$ region, where $m_1$ is the ZAMS mass of the primary, defined as the initially most massive component of the system.  Because these ranges are not uniformly covered by all source surveys, not all fits apply to them in full; we indicated where we have extrapolated or otherwise deviated from the original fits. With respect to $m_1$, we always extend the distributions to $0.8\leq m_1/\Msun \leq 150$; while we keep to the $0.1<q<1.0$ and $0.2<\log \lrp{P/1\text{ d}}<8.0$ region. For the sake of comparing the two models, we extend the Invariant distributions to cover the same ranges as the Varying distributions.

Below, we discuss distributions for $q$, $P$, $e$ and also multiplicity fractions in turn. $\log P$ is used as shorthand for $\log\lrp{P/1\text{ d}}$. Besides the parameter distributions themselves, \citetalias{MoeDistefano2017} also fit $1\sigma$ uncertainties for all parameters; we, however, do not include these as our aim is to capture the average behavior. Finally, it is fundamental to note that these distributions describe multiple systems of arbitrary order (singles, binaries, triples...), as further discussed in Secs. \ref{sec2sub:multiplicity} and \ref{sec4sub:multiplicity}. For this reason we use the term \textit{pair} to mean a pair made up of the system primary and one of its $\ncp\geq1$ companions; while we reserve \textit{binary} for multiple systems with $\ncp=1$ companion, specifically. It is also important to keep this in mind when comparing the Varying distributions to the Invariant ones, as the latter are supposed to reflect properties of actual binary systems only; while even if the Varying models could still lead to distributions similar to the Invariant in the case of binaries, it is in principle possible to separate the contributions of different-order multiples, which cannot be done reliably at the moment. This issue is discussed in Sec. \ref{sec2sub:multiplicity} and further investigated in Sec. \ref{sec4sub:multiplicity}.

\subsubsection{Mass ratio distribution}
\label{sec2sub:massratio}

\begin{figure}
    \centering
    \includegraphics[width=\columnwidth]{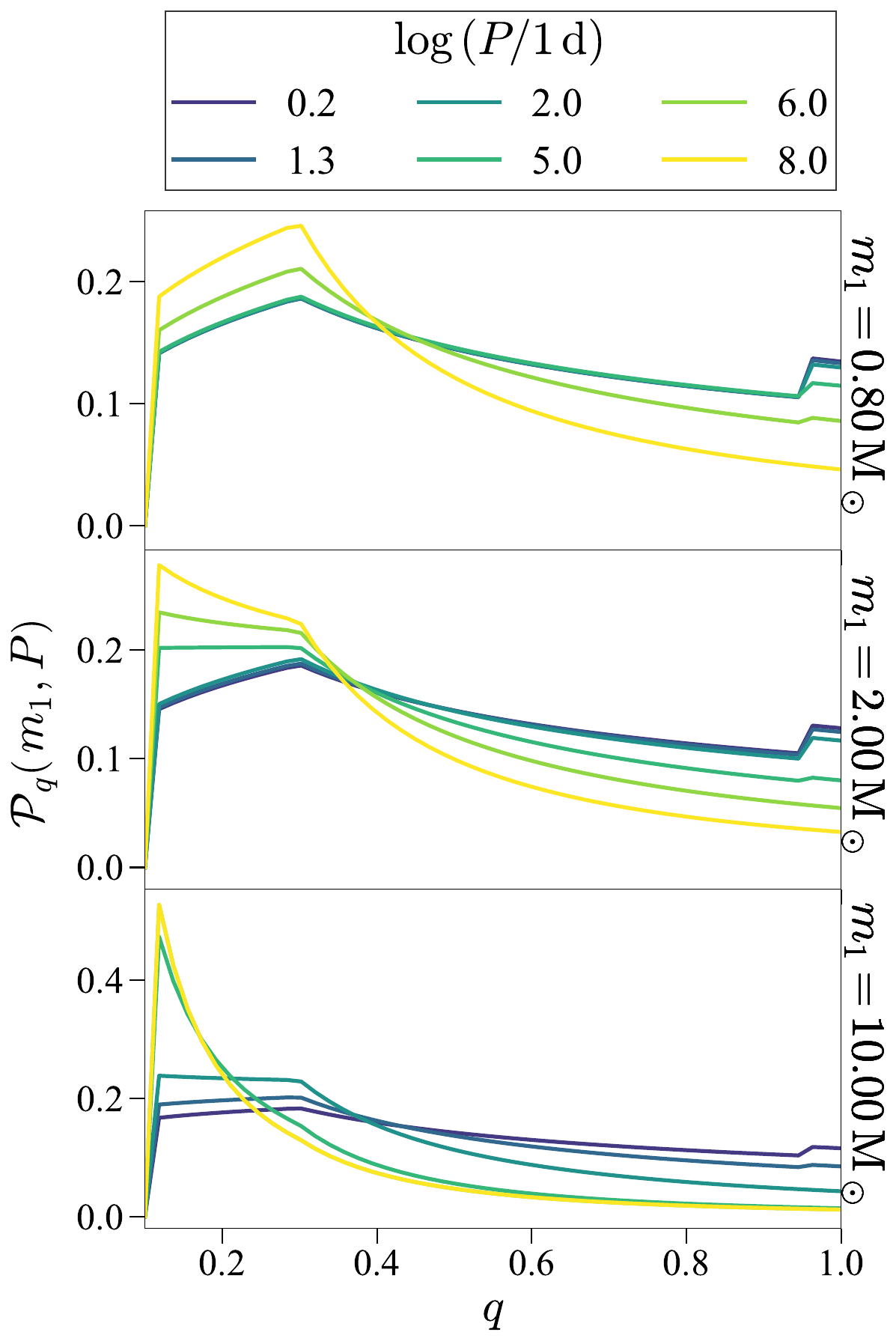}
    \caption{Mass ratio distribution for $0.1\leq q\leq1.0$, for representative $\logP$ between $0.2$ and $8$, and primary masses $m_1$ between $0.8$ and $10\,\Msun$. For the lightest primaries (upper panel) there is a peak at $q=0.3$ regardless of the orbital period, and there is a non-zero $\ftwin$ for $\logP\lesssim6$ (upper panel). For masses around $m_1=2\,\Msun$ (middle panel), the distributions peaks at $q=0.3$ for $\logP\lesssim5$, and at $q=0.1$ otherwise; $\ftwin$ is non-zero for $\logP\lesssim5$. For the most massive primaries (lower panel), the distribution peaks at $q=0.1$ for $\logP\gtrsim5$, otherwise it plateaus below $q=0.3$ and smoothly drops for greater $q$; a small non-zero $\ftwin$ appears for $\logP\gtrsim1.3$. In all cases the distribution approaches being uniform as the orbital period decreases, in particularly for more massive primaries. The only significant variation in the shape of the distribution beyond $m_1=10\,\Msun$ is that $\ftwin$ continues to decrease for the closest orbits. All curves are normalized so that the area under each is unity.}
    \label{fig:md17_qdistr}
\end{figure}

For the mass ratio distribution, $\mathcal{P}_q$, \citetalias{MoeDistefano2017} fit a two-part power law, with a \textit{small $q$} range up to $0.3$, and a \textit{large $q$} range beyond; a \textit{twin excess} is also added as an extra multiplicative factor to the distribution above $q=0.95$. All distribution parameters are functions of $m_1$ and $\logP$. As illustrated in Fig. \ref{fig:md17_qdistr}, as $m_1$ increases, the distribution peak shifts from $q=0.3$ to $q=0.1$, the latter generally being the case for compact object (CO) progenitors ($m_1\gtrsim8\,\Msun$), for which the twin excess is not significant; increasing orbital periods shift the distribution towards asymmetric pairs, i.e., the $q=0.1$ peak. For a fixed orbital period, the shape of the distribution remains essentially invariant between $10$ and $40\,\Msun$; thus our extrapolation up to $150\,\Msun$ keeps the same form as in that range.

Broadly speaking, the closer a pair is, the more symmetric its component masses tend to be. However, the more massive a primary is, the less likely it is to have a companion of similar mass, regardless of their separation. This behavior is proposed by the authors to reflect two different mechanisms of pair formation: molecular core/filament fragmentation, predominantly responsible for pairs with separations $a\gtrsim200\,\mathrm{au}$; and disk fragmentation, predominantly responsible for $a\lesssim200\,\mathrm{au}$ pairs. Pairs formed from disk fragmentation undergo competitive accretion from the disk, thus leading to more symmetric masses; as disk lifetimes tend to decrease with increasing protostar mass, this may also be the reason for massive primaries tending towards less symmetric pairings, even in close orbits.

\subsubsection{Eccentricity distribution}
\label{sec2sub:eccentricity}

\begin{figure}
    \centering
    \includegraphics[width=\columnwidth]{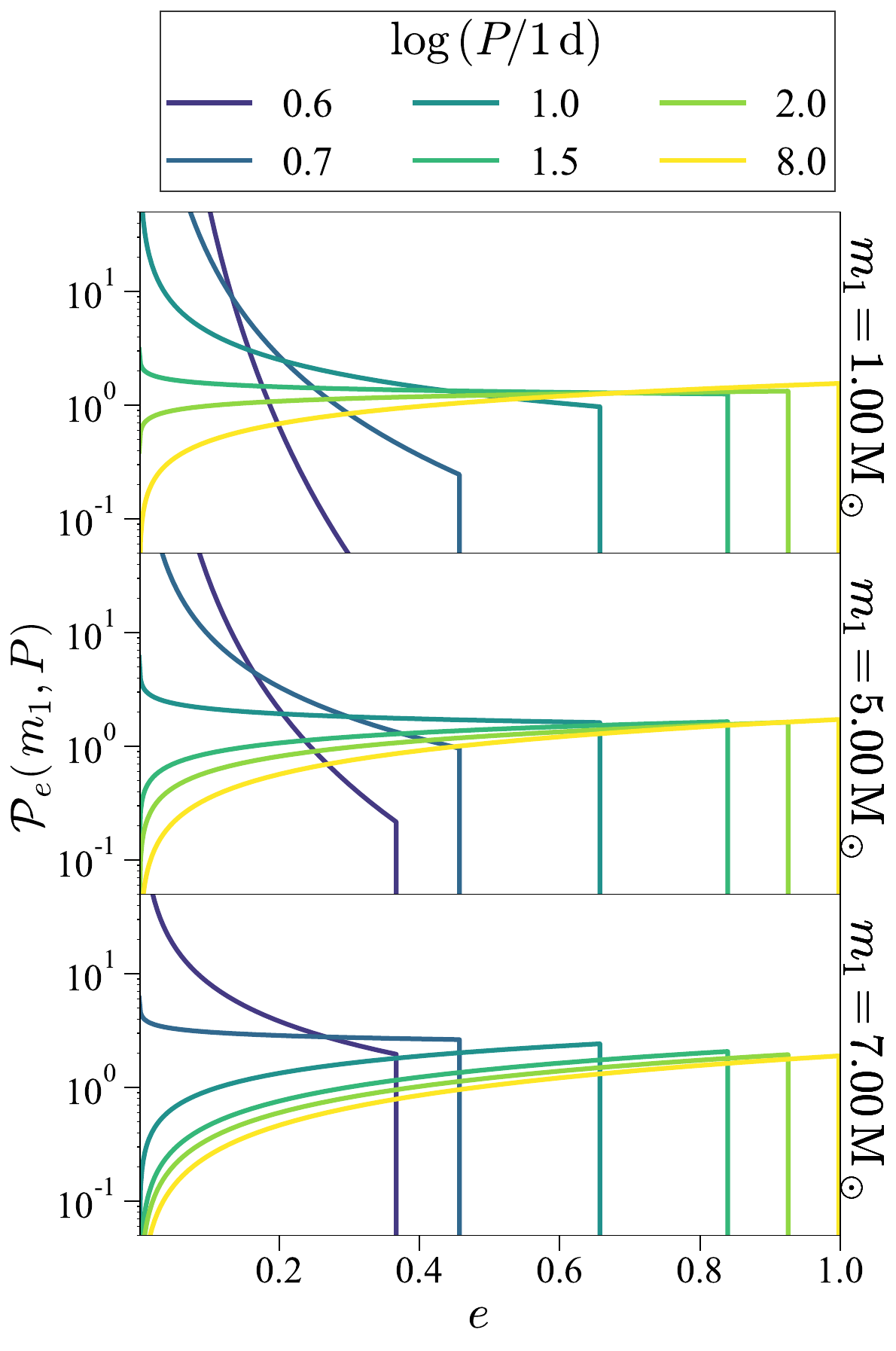}
    \caption{Eccentricity distribution for $0< e\leq1$, for representative orbital periods $P$ between $10^{0.6}$ and $10^8\text{ d}$, and primary masses $m_1$ between $1$ and $7\,\Msun$. For the lightest primaries (upper panel), orbits with $\logP\lesssim1.5$ strongly favor $e=0$; around $m_1=5\,\Msun$  (middle panel), only orbits with $\logP\lesssim1$ strongly favor $e=0$; and at $m_1=7\,\Msun$ (lower panel), only those with $\logP\lesssim0.7$. The distribution shifts very rapidly in the immediate neighborhood of these thresholds. Whenever the distribution doesn't peak sharply at $e=0$, it is almost uniform. The shape does not change, for a given period, beyond $m_1=7\,\Msun$. The breaks in the curves towards higher eccentricities indicate $e_{\max}$, which approaches (but never reaches) $1$ for increasing periods. All orbits with $\logP<0.5$ are assumed to be circularized. All curves are normalized so that the area under each is unity.}
    \label{fig:md17_edistr}
\end{figure}

For the eccentricity distribution, $\mathcal{P}_e$, \citetalias{MoeDistefano2017} fit a simple power law between $e=0$ and $e_\mathrm{max}$. The exponent is fitted as dependent on both $m_1$ and $P$; while $e_\mathrm{max}$ is a function of $P$ imposing that the pair of components have Roche lobe filling factors $\lesssim70\%$ at periastron, in order to avoid interaction. All $\logP\leq0.5$ orbits are assumed to be circularized at ZAMS. Some example cases are shown in Fig. \ref{fig:md17_edistr}. For the least massive primaries, the distribution heavily favors $e=0$ for $\logP\lesssim1.5$ and approaches a uniform one for $\logP\gtrsim1.5$. As $m_1$ increases, the distribution becomes more uniform for all but the closest orbits, and remains essentially invariant beyond $m_1=7\,\Msun$, for which it still strongly favors $e=0$ if $\logP\lesssim0.7$, but is close to uniform otherwise. The maximum eccentricity becomes greater than $0.9$ for $\logP\approx2$, and continues to approach $1$ as the period increases. Similarly to $\mathcal{P}_q$, for a fixed orbital period $\mathcal{P}_e$ is found to be essentially invariant between $7\,\Msun$ and $40\,\Msun$, thus we keep this shape in our extrapolation up to $150\,\Msun$.

The power law exponents were fitted only up to $\logP=6$ (for $m_1<3\,\Msun$) and $\logP=5$ (for $m_1>7\,\Msun$), as the nature of the eccentricity distribution for longer periods is currently speculative at best. \citetalias{MoeDistefano2017} suggest that, for solar-type primaries, the value $\approx0.5$ to which the slope saturates close to $\logP=6$ might extend to the widest separations, which we assume to be true. The same cannot be said of early-type primaries, however, as the $\approx0.8$ to which it saturates at $\logP=5$ cannot be valid for the same $0<e<e_{\max}\approx1$ range for increasingly wider orbits. This is because the majority of wide companions to O- and B-type primaries are in higher order systems where additional stability criteria must determine $e_{\max}$. While $\eta\approx0.8$ might continue to higher $P$, $e_{\max}$ must, at least, change.

As stated above, for the purposes of BPS we are only interested in binaries and (in at least some cases) the inner binaries of higher-order hierarchical systems. The first are not affected by the destabilizing influence of additional companions, while the second are more likely intermediate to low-period systems, such that they are not heavily affected by the above uncertainties. This alleviates the error induced by the extrapolation of the eccentricity distribution.\footnote{One caveat is that we still might draw second and further companions to track how much mass is formed in higher-order multiples, and this requires drawing wide orbits, which will push the extrapolation. This is not an issue as long as only the mass is of interest, not the evolution of higher-order multiples. See Secs. \ref{sec2sub:multiplicity} and \ref{sec4sub:multiplicity}.}

The increased probability of closer pairs to be circularized emerges naturally as a consequence of stronger tidal forces. For longer periods, the distribution approaches a thermal distribution (slope $\approx1$), which the authors propose indicates a greater role of dynamical interactions in the formation of these systems.

\subsubsection{Orbital period distribution}
\label{sec2sub:period}

The characterization of the orbital period is less straightforward than that of the mass ratio and eccentricity. First, the quantity being modeled is not a probability density like $\mathcal{P}_q$ and $\mathcal{P}_e$, but the \textit{companion frequency},

\begin{equation}
    \label{sec2eq:fcomp_q03}
    f_{\log P;q\geq q_\mathrm{min}}(m_1) = \frac{\d N_\mathrm{cp}}{\d N_1\d\log P}(m_1),  
\end{equation}

\noindent which expresses the number of companions, $N_\mathrm{cp}$, per primary of a given mass ($N_1$, $m_1$), per decade of orbital period, in pairs with mass ratio greater than some $q_{\min}$. While we will have more to discuss about the interpretation and implementation quantity in Sec. \ref{sec2sub:multiplicity}, here we focus on the shape of the distribution. In order to build a continuous model, \citetalias{MoeDistefano2017} find from their data the companion frequency as a function of $m_1$ in $\logP<1$, at $\logP=2.7$, and at $\logP=5.5$; then, $f_{\log P;q\geq q_\mathrm{min}}$ is fitted, for a fixed $m_1$, in the entire $0.2<\logP<8$ range as a piecewise log-linear function of the orbital period, with a exponential tail beyond $\logP=5.5$. 

Because their orbital period data is only complete down to $q=0.3$, their fit only applies to $q>0.3$ pairs. In order to extend the description down to $q=0.1$, \citetalias{MoeDistefano2017} suggest extrapolating from $\mathcal{P}_q$, which does extend down to $q=0.1$, a procedure which we followed. By integrating $\mathcal{P}_q$ in the appropriate ranges, for each $\lrp{m_1,P}$ it is possible to compute the fraction of $q>0.1$ pairs in relation to that of $q>0.3$ pairs. Then we assume that,

\begin{align}
    f_{\logP;q\geq0.1}&\lrp{m_1,P} = \nonumber\\
    &\frac{\int_{0.1}^{1.0}\;\prob_q\lrp{m_1,P}\;\d q}{\int_{0.3}^{1.0}\;\prob_q\lrp{m_1,P}\;\d q} f_{\logP;q\geq0.3}\lrp{m_1,P}. \label{sec2eq:fcomp_q01}
\end{align}

Unlike $\mathcal{P}_q$ and $\mathcal{P}_e$, $f_{\logP;q\geq0.1}\lrp{m_1,P}$ does vary up to and beyond $m_1=40\,\Msun$. We have chosen to extrapolate the fit by \citetalias{MoeDistefano2017} and assume that the distribution continues to evolve in the same way up to $m_1=150\,\Msun$; an alternative is to assume that it keeps its shape at $m_1=40\,\Msun$ beyond it, as done by \citet{Klencki2018}. 

\begin{figure}
    \centering
    \includegraphics[width=\columnwidth]{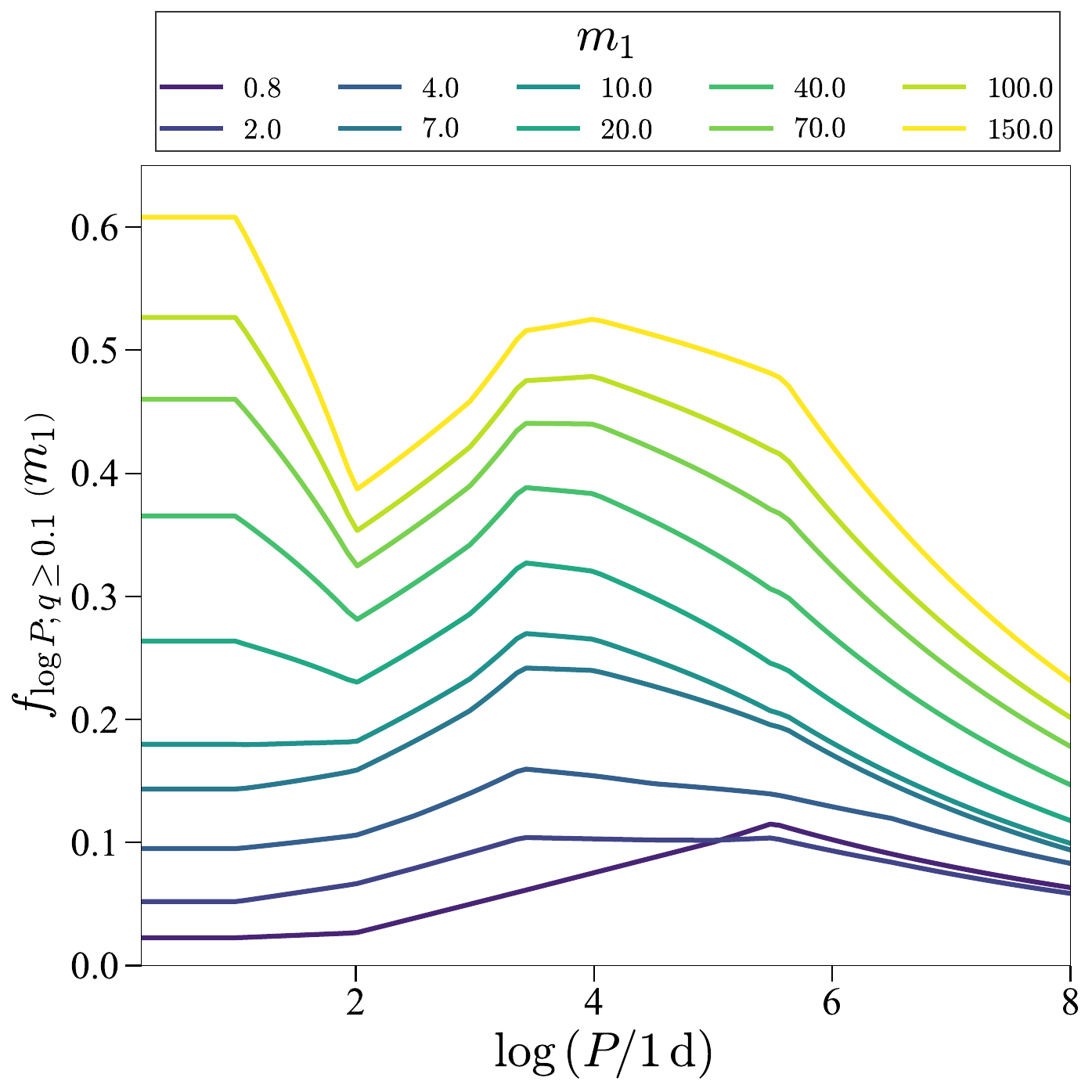}
    \caption{Companion frequency as a function of orbital period for $q\geq0.1$ pairs over primary mass $m_1$ and orbital period $P$, with $m_1$ between $0.08$ and $150\,\Msun$, and $\log\lrp{P/\text{1d}}$ between $0.2$ and $8$. Solar-type primaries approach a log-uniform distribution over $P$, but show a peak at $\logP=5.5$ thought to be connected to core/filament fragmentation. Early-type primaries shift to a lower $\logP=3.4$ peak, proposed to originate with disk fragmentation. The $\logP$ plateau which develops for the most massive primaries might reflect a channel of dynamical evolution of pairs stemming from disk fragmentation. In this scenario, the overall increase in the multiplicity fraction with mass is a natural consequence of the increased probability of disk fragmentation for massive protostars. Although the companion frequency does not distinguish between multiple order (binary, triple...), it can be read as a companion orbital period probability distribution.}
    \label{fig:md17_compfreq}
\end{figure}

The companion frequency is shown as a function of orbital period, for different $m_1$, in Fig. \ref{fig:md17_compfreq}. The breaks in the piecewise fit lead to the sharp features at $\logP=1,2,3.4$ and $5.5$, with the lower features increasing in importance for increasing $m_1$. For early-type primaries, the companion frequency is dominated by a peak near $\logP=4$ and a plateau in $\logP=0.2-1$. Both of these features have about the same weight at $m_1=40\,\Msun$, but, due to our choice of extrapolation, the plateau becomes increasingly dominant for higher masses. The continued increase in the absolute values of the companion frequency with mass also indicates that the average number of companions per primary monotonically increases with $m_1$ at all orbital periods; the only exception is the range between $0.8$ and $2\,\Msun$ for $\logP\gtrsim5$.

The shape of the distribution is suggested to capture the competing effects of the molecular core/filament fragmentation and disk fragmentation formation channels described in Sec. \ref{sec2sub:massratio}. Because massive disks are much more prone to fragmentation, this channel would play a dominant role in massive pair formation, thus leading to the shift towards the $\logP=3.4$ peak; the less massive the primary, the less likely disk fragmentation is, and the more dominated the distribution becomes by the $\logP=5.5$ "core/filament fragmentation peak". The increased likelihood of disk fragmentation around massive protostars would also explain the overall increase of the companion frequency for massive primaries. The bimodality of the distribution for massive primaries, with the plateau below $\logP=1$, is suggested to indicate a different formation mechanism for these close binaries. \citetalias{MoeDistefano2017} suggest this to be be a dynamical evolution, still in the pre- ZAMS phase, of a wider multiple system initially formed from disk fragmentation, with an outer tertiary in a particular configuration. In this scenario, the choice between extrapolating the evolution of the distribution beyond $40\,\Msun$ (as performed in our work) or keeping it invariant in that range becomes a matter of whether such a dynamical evolution channel becomes more likely or not with the increasing multiplicity of primary masses. Although our choice implies it does, this is not necessarily the case given that this channel would likely depend on particular orbital configurations of outer companions.

\subsubsection{Multiplicity fractions}
\label{sec2sub:multiplicity}

In order to normalize our results to physical binary populations, we must know the total \textit{star-forming mass} ($\Msf$) to which each binary sample corresponds, i.e., the total stellar mass in binaries, plus that in isolated stars and in higher-order multiples. Knowledge of this quantity allows us to compute certain properties, such as merger rates \textit{per stellar mass}, from which we can find the total merger rate for a physical population of known stellar mass.

The first important quantity necessary for recovering $\Msf$ is the binary fraction: the fraction of \textit{primaries} with $\ncp=1$ companion. For consistency with the mass ratio distribution by \citet{SanaMassRatio}, this has often been fixed at $0.7$ for CO progenitor primaries \citep[e.g.,][]{Neijssel2019,07binaryfrac2020zevin,07binaryfrac2021bavera,07binaryfrac2021roman-garza,vanSon2022,07binaryfrac2024dorozsmai}. When extrapolating to initial orbital periods greater then the original upper limit of $\logP=3.5$ in \citet{SanaMassRatio}, \citet{deMink2015} find that a binary fraction of $1$ is more adequate. This value has also been widely used \citep[e.g.,][]{1binaryfrac2018chruslinska,1binaryfrac2018kruckow,1binaryfrac2020belczynski,1binaryfrac2021shao,broekgaarden2021,broekgaarden2022}.

In the case of the Varying distributions, the binary fraction is bound to the companion frequency, which already contains information about the number of companions per primary for multiples or even arbitrary order, as discussed above. If we allow for primaries with up to $\ncp^\mathrm{max}$ companions, then the \textit{multiplicity fractions} $\mathcal{F}_{\ncp;q\geq0.1}\lrp{m_1}$ are related to the companion frequency by

\begin{equation}
     f_{\mathrm{mult};q\geq0.1}\lrp{m_1} = \sum_{\ncp=0}^{\ncp^\mathrm{max}}\ncp\mathcal{F}_{\ncp;q\geq0.1}\lrp{m_1},
     \label{sec2eq:fmult_multfracs}
\end{equation}

\noindent where $\mathcal{F}_{\ncp;q\geq0.1}\lrp{m_1}$ is the fraction of primaries with $m_1$ that have $\ncp$ companions with $q\geq0.1$; and $f_{\mathrm{mult};q\geq0.1}\lrp{m_1}$ is the \textit{multiplicity frequency}, obtained by integrating the companion frequency (equations \eqref{sec2eq:fcomp_q01} and \eqref{sec2eq:fcomp_q03}) over the full orbital period range, which expresses the average number of companions per primary of mass $m_1$. The fraction of isolated primaries is simply 

\begin{equation}
    \mathcal{F}_{\ncp=0;q\geq0.1}\lrp{m_1}=1-\sum_{\ncp=1}^{\ncp^\mathrm{max}}\mathcal{F}_{\ncp;q\geq0.1}\lrp{m_1}.
    \label{sec2eq:isolated_frac}
\end{equation}

Although integration of $f_{\mathrm{\logP;q\geq0.1}}$ leads to $f_{\mathrm{mult};q\geq0.1}$, we cannot solve equation \ref{sec2eq:fmult_multfracs} without additional information. Here, we follow \citetalias{MoeDistefano2017} in assuming that the probability distribution for $\ncp$, $\mathcal{P}_{\ncp}$, for \textit{all} primaries assumes a Poissonian profile truncated at $[0,\ncp^\mathrm{max}]$, as has been observed for solar-type binaries. Then,

\begin{equation}
    \label{eq:poisson}
    \prob_{\ncp}\lrp{m_1} = \frac{\lrs{\lambda\lrp{m_1}}^{\ncp}}{\ncp!}e^{-\lambda\lrp{m_1}}, \quad \ncp\leq\ncp^\mathrm{max},
\end{equation}

\noindent and the expectation value of $\ncp$, $\lambda\lrp{m_1}$, is found by imposing $\mathcal{P}_{\ncp=n}\lrp{m_1=m}=\mathcal{F}_{\ncp=n;q\geq0.1}\lrp{m_1=m}$ and equation \eqref{sec2eq:fmult_multfracs}. 

\begin{figure}
    \centering
    \includegraphics[width=\columnwidth]{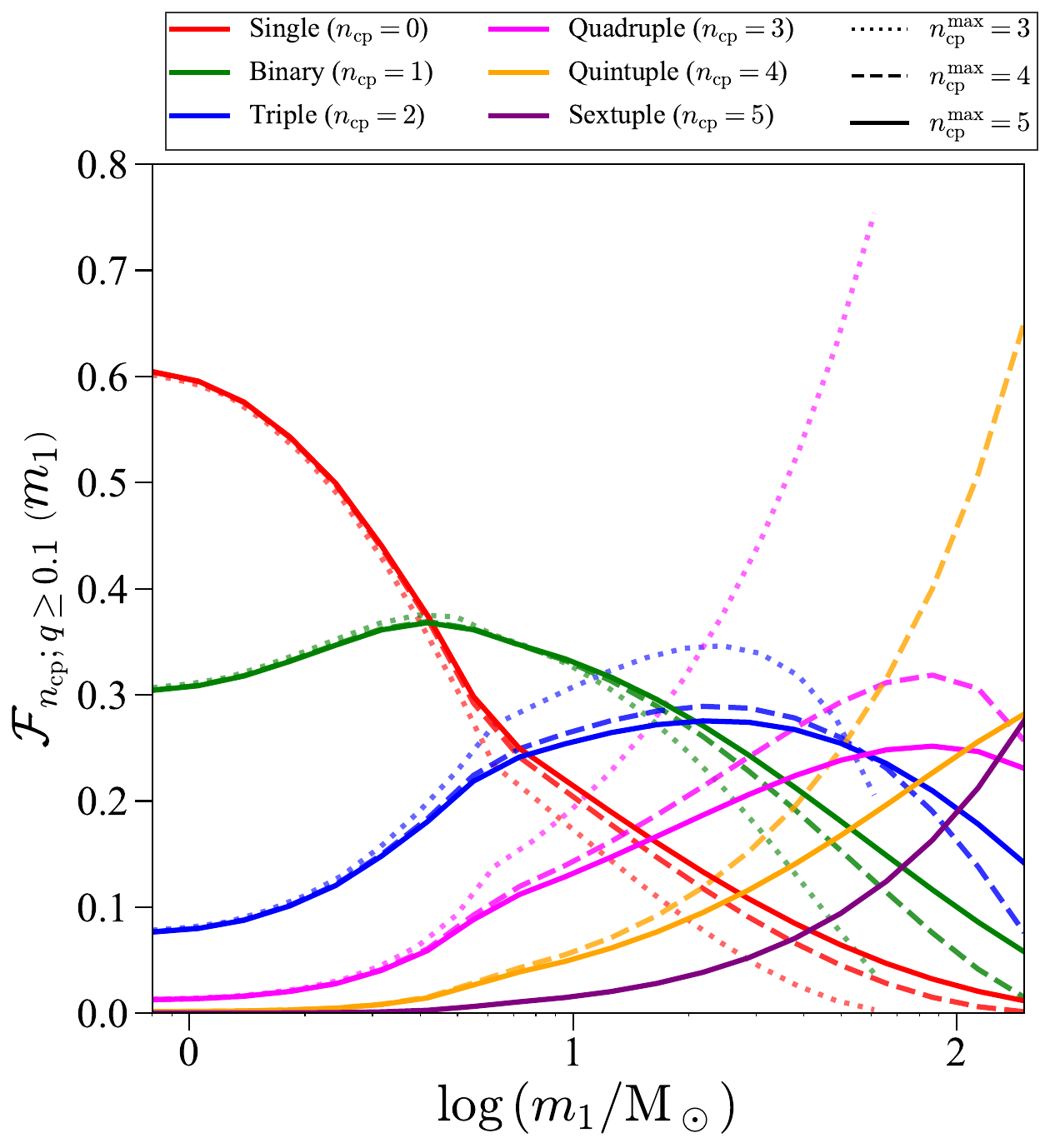}
    \caption{Multiplicity fractions $\mathcal{F}_{\ncp;q\geq1}\lrp{m_1}$ as a function of $m_1$ between $0.8$ and $150\,\Msun$, for the cases $\ncp^{\max}=3,4$ and $5$ (solid, dashed and dotted lines, respectively), in the AM model. For $\ncp^\mathrm{max}$, equation \ref{sec2eq:fmult_multfracs} has a solution only up to $\sim60\,\Msun$. As $\ncp^\mathrm{max}$ is raised, the curves converge to a particular shape; we consider that $\ncp^\mathrm{max}=4$, besides being the minimum necessary for the solution to exist up to $m_1=150\,\Msun$, approaches well enough this limit shape and does not miss any strong behavior of the multiplicity fractions.}
    \label{sec2fig:allmult_frac}
\end{figure}

Because $f_{\mathrm{mult};q\geq0.1}$ grows monotonically with mass, the minimum required $\ncp^\mathrm{max}$ for equation \eqref{sec2eq:fmult_multfracs} to have a solution in a given $[m_\mathrm{min},m_\mathrm{max}]$ range grows with $m_\mathrm{max}$. We find that a solution up to our maximum $m_1=150\,\Msun$ requires at least $\ncp^\mathrm{max}=4$, because $f_{\mathrm{mult};q\geq0.1}>3$ for $m_1\gtrsim60\,\Msun$; we adopt this value as our maximum companion number, while \citetalias{MoeDistefano2017} had adopted $\ncp^\mathrm{max}=3$ in order to solve the equation up to $m_1=40\,\Msun$. In Fig. \ref{sec2fig:allmult_frac}, we show the resulting multiplicity fractions from solving equation \eqref{sec2eq:fmult_multfracs} for $m_1$ between $0.8$ and $150\,\Msun$, for $\ncp^\mathrm{max}=3,4$ and $5$. When varying $\ncp^\mathrm{max}$ up to $4$, the only significant change in the shape of the curves is for the quadruple fraction, which increases monotonically if $\ncp^\mathrm{max}<4$ and displays a peak otherwise. The $\ncp^\mathrm{max}=5$ case is included in Fig. \ref{sec2fig:allmult_frac} in order to demonstrate that the curves converge to a particular shape as $\ncp^\mathrm{max}$ is raised, and that we can consider the $\ncp^\mathrm{max}=4$ case to be close enough to the limit shape and not to miss any key behavior (such as a peak). The multiplicity fractions for up to $\ncp=2$ behave similarly in the solutions for $\ncp^\mathrm{max}= \{3,4\}$ in the entire range, while the quadruple fraction ($\ncp=3$) has a peak beyond $40\,\Msun$, instead of growing indefinitely, when changing from $\ncp^\mathrm{max}=3$ to $4$; for $\ncp^\mathrm{max}=4$, the fraction of quintuples ($\ncp=4$) grows up to $\sim0.2$ at $40\,\Msun$.

The necessity of admitting the presence of higher-order multiples in the population in order to obtain a consistent solution for the binary fraction holds some implications worth minding when performing BPS. Many phenomena that arise in binary evolution --- such as, e.g., pulsar recycling and accretion-induced collapse, in addition to compact object mergers --- should arise also in the evolution of higher-order multiples, and so a comprehensive study ought to account for populations of \textit{multiples} more generally. While at present BPS is still often employed on its own to estimate the rates and properties of these events, we may still more carefully consider exactly \textit{which} population should be contributing to, for example, estimated merger rates, as the fractions in Fig. \ref{sec2fig:allmult_frac} allow distinguishing the pure binary fraction from the fraction of systems with a least one companion. A choice can then be made, within the limitations of BPS, to evolve solely binaries, or also include those inner binaries of higher-order multiples that can be expected to evolve like binaries. This is important, because under the right conditions the presence of a further companion can drastically alter the evolution of the inner binary, through mechanisms such as mass transfer \citep[e.g.,][]{tripleinteraction2014devries,tripleinteraction2020comerford,tripleinteraction2020distefano,tripleinteraction2022hamers}; secular evolution of the inner binary (on a timescale longer than the orbital periods), as with Lidov-Kozai (LK) cycles \citep[e.g.,][]{toonen2020triples,stegmann2022triples,tripleinteraction2023kummer} or eccentric LK cycles \citep[see the review by][]{naoz2016eccKLcycles}, which can increase the efficiency of gravitational radiation by driving the inner binary to high eccentricities; or even semisecular evolution of the inner binary, in which the tertiary-induced perturbations on the eccentricity of the inner binary occur in a timescale equal to or shorter then the orbital periods \citep[e.g.,][]{Antonini2014semisecular,antognini2014semisecular}. An analogous role can be played by LK cycles in quadruple systems as well \citep[e.g.,][]{Fragione2020hierarch_spin,quadrupleev2020safarzadeh,Vynatheya2022hierach}. For the particular case of triples, we point to the example of \citet{Klencki2018}, who account for the triple fraction when sampling systems from the \citetalias{MoeDistefano2017} correlated distributions in order to study BCMs.

\begin{figure}
    \centering
    \includegraphics[width=\columnwidth]{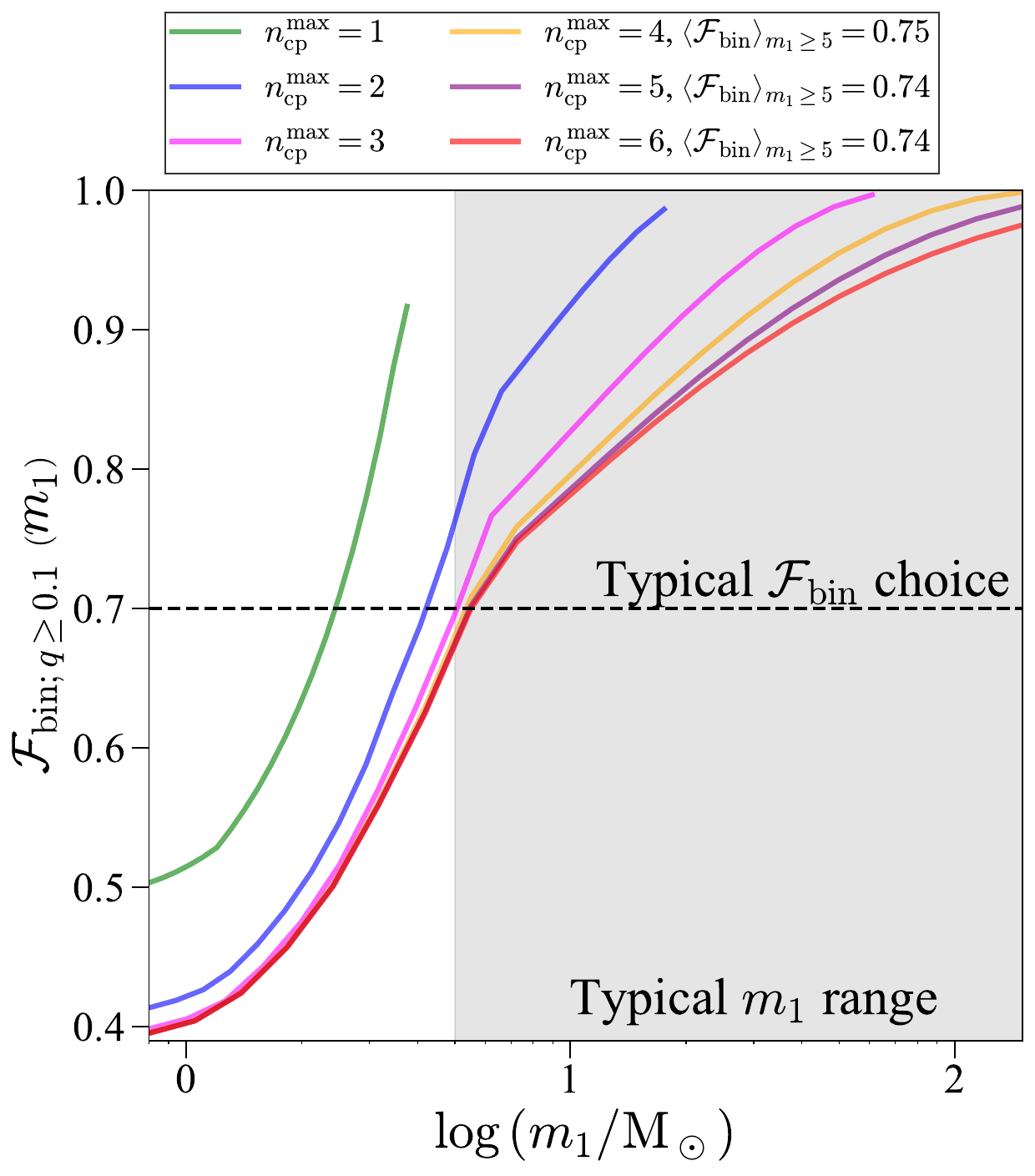}
    \caption{The binary fraction, $\mathcal{F}_{\mathrm{bin};q\geq0.1}$, as a function of $m_1$, between $0.08$ and $150\,\Msun$, computed in the OB model, for $\ncp^\mathrm{max}$ from $1$ to $6$. The greater $\ncp^\mathrm{max}$ is, the higher is the upper limit of the mass range for which equation \eqref{sec2eq:fmult_multfracs} has a solution, and for which we can compute $\mathcal{F}_\mathrm{bin;q\geq0.1}$. The gray area indicates the $[5\,\Msun,150\,\Msun]$ range from which $m_1$ is often sampled in BPS studies concerned with BCO populations, and the horizontal dashed line the typical $\mathcal{F}_\mathrm{bin}=0.7$ chosen for this region, for consistency with \citet{SanaMassRatio}. In the legend, for the cases in which we can compute the binary fraction up to $m_1=150\,\Msun$, we show the Salpeter IMF-weighted average $\mathcal{F}_{\mathrm{bin};q\geq0.1}$ in the $[5\,\Msun,150\,\Msun]$ range. This average trends towards $0.74$, consistent with \citet{SanaMassRatio}.}
    \label{sec2fig:onlybin_frac}
\end{figure}

Regardless of whether inner binaries of higher-order multiples are later evolved or not, we call the model in Fig. \ref{sec2fig:allmult_frac} the \textit{AM (All Multiples)} model. It has one clear inconsistency with typical assumptions in BPS: a binary fraction which at its peak only reaches up to $0.4$, in contrast to the $0.7$ based on \citet{SanaMassRatio}. This decrease is a natural consequence of taking multiples into account and treating higher-order multiples separately from binaries. We can see this more clearly by defining

\begin{equation}
    \label{sec2eq:binfrac}
    \mathcal{F}_{\mathrm{bin};q\geq0.1}\lrp{m_1} = \sum_{\ncp=1}^{\ncp^\mathrm{max}}\mathcal{F}_{\ncp;q\geq0.1\lrp{m_1}},
\end{equation}

\noindent i.e., by forcing all multiple systems to be binaries. This results in the binary fraction as a function of $m_1$ shown in Fig. \ref{sec2fig:onlybin_frac}, for $\ncp^\mathrm{max}$ up to $6$. At $m_1=5\,\Msun$ we have $\mathcal{F}_\mathrm{bin;q\geq0.1}\approx$ converging towards $0.7$, and approaching $1$ at $m_1=150\,\Msun$. For the cases where equation \eqref{sec2eq:fmult_multfracs} has a solution up to $150\,\Msun$ ($\ncp^\mathrm{max}\geq4$), we find that the \citet{Salpeter1955} IMF-weighted average $\mathcal{F}_\mathrm{bin;q\geq0.1}$ converges to $0.74$, which is consistent with the $\mathcal{F}_\mathrm{bin}=0.69\pm0.09$ found by \citet{SanaMassRatio}. We thus define the \textit{Only Binaries (OB)} model as the one in which we consider only isolated stars and binaries, according to $\mathcal{F}_\mathrm{bin;q\geq0.1}$ computed with $\ncp^\mathrm{max}=4$ (equation \ref{sec2eq:binfrac}).

The OB approach is closer to the majority of BPS works, while also offering a refinement of the normalization process by accounting for the growing of multiplicity with mass. Finally, we highlight that, in terms of BPS, the OB model is not equivalent to taking the AM model and evolving all inner binaries; the latter case induces a bias towards short orbital periods for inner binaries of higher-order multiples which is not present in the former (see Sec. \ref{sec:3sampling}).

\subsection{Environmental conditions}
\label{sec2sub:environment}

The last pieces of information required are a) the metallicity of the environment in which the formation of a given population took place, which partially determines the IMF, in the Varying model, and affects massive stellar evolution; b) the SFR of that same environment, which determines the Varying IMF along with the metallicity; and c) the formation redshift, which defines the age of the population.

For these \textit{environmental conditions}, we base our work on \citet{Chruslinska2019} (\citetalias{Chruslinska2019} hereon) and \citet{Chruslinska2020} (\citetalias{Chruslinska2020} hereon). In order to study the evolution of stellar metallicity over time, \citet{Chruslinska2019} model the cosmic star formation history (cSFH), and its uncertainties, with a collection of models for the galaxy stellar mass function (GSMF), the star formation-mass relation (SFMR) and the mass-metallicity relation (MZR). All three are empirical relations that describe the average properties of star-forming galaxies at different redshift, fitted on data from up to $z\sim9$ in the first two cases, and $z\sim3.5$ on the last. All relations are extrapolated up to $z=10$, which is assumed to mark the start of star-formation.

The GSMF, $\Phi\lrp{M_\ast|z}$, gives us the number density of galaxies as a function of the galaxy stellar mass, $M_\ast$, at each redshift. The fit by \citetalias{Chruslinska2019} is built from data of 13 different sources, each covering a different redshift bin, and thus represents an average over them, as they do not always individually agree. The main uncertainty is with regard to lower-mass galaxies, hence dimmer and harder to observe; thus they include two models of the GSMF  that differ with regard to the slope at low masses: the Fixed Slope model, and the redshift-dependent Varying Slope model.

\begin{figure}
    \centering
    \includegraphics[width=\columnwidth]{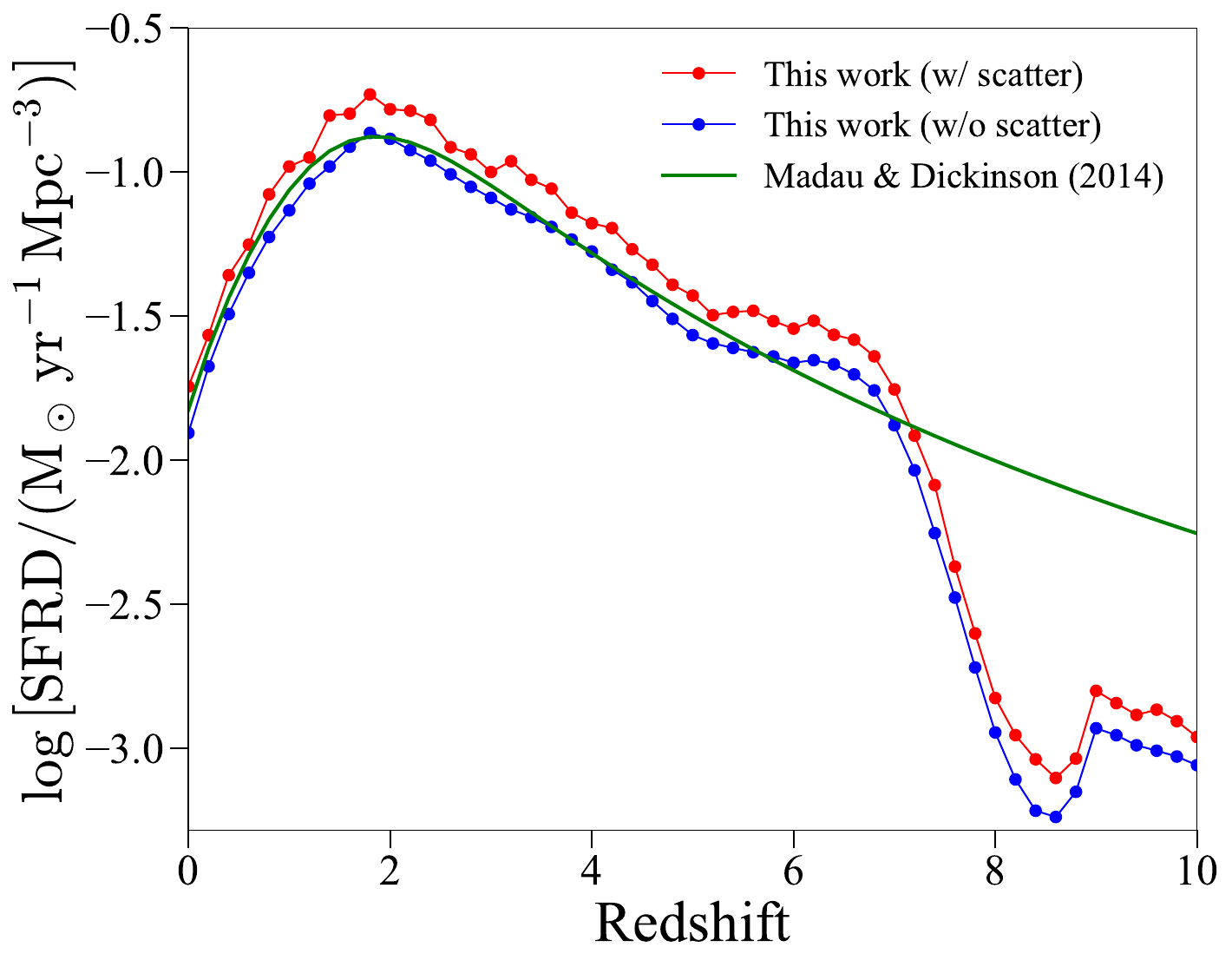}
    \caption{SFRD as a function of redshift from \citet{SFHMadau} (solid green line), compared with that obtained from the Moderate Metallicity model described in Sec. \ref{sec2sub:environment} both when ignoring (blue solid line+circles) and when accounting for (red solid line+circles) scatter in the metallicity and SFR. The blue and red curves are calculated following steps analogous to those in \citet{Chruslinska2019}, described in Sec. \ref{sec3sub:gal_sampling}.}
    \label{sec3fig:madau}
\end{figure}

\begin{figure*}
    \centering
    \includegraphics[width=\textwidth]{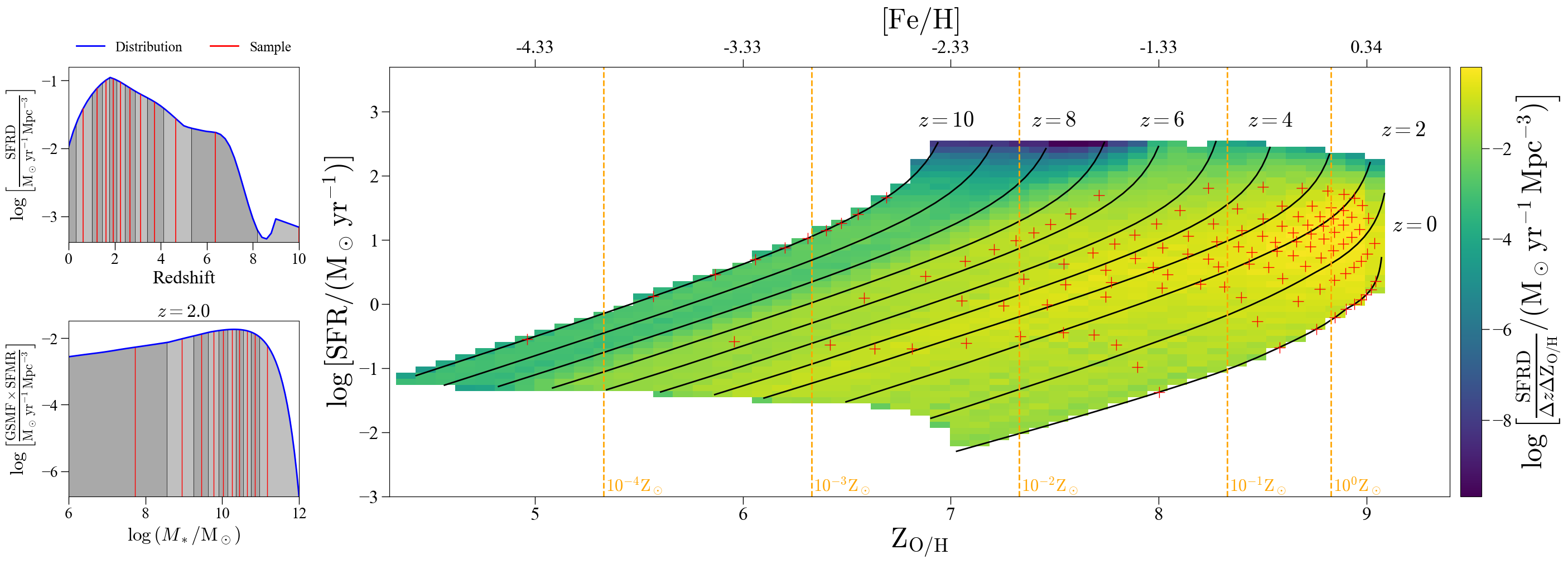}
    \caption{\textit{Right:} Star-formation rate density (SFRD, background shaded plot) distribution on the SFR-metallicity plane, based on the the moderate metallicity model of \citet{Chruslinska2020} for the GSMF, SFMR and MZR without scatter, for a Varying IMF assumption. The background SFRD was generated from a sample of $250,000$ galaxies. The solid black lines indicates constant redshift, while the dashed orange lines indicate metallicities in units of solar metallicity. The red crosses show the result of sampling without scattering of $10$ redshifts, plus manually set boundary redshifts $0$ and $10$; and of 10 metallicities per redshift. The boundaries of the shaded plot corresponds to the region covered by galaxies with stellar mass between $10^6$ and $10^{12}\,\Msun$. The SFRD is computed within 2D bins of fixed widths $\Delta \log\SFR=0.095$ and $\Delta\ZOH=0.095$. Sampling is weighted by the SFRD, and examples are shown on the left. \textit{Top-left:} sampling without scattering of 10 redshifts. The SFRD (blue line) is divided into 10 quantiles (gray bands), and each one is represented by its SFRD-weighted average redshift (vertical thick red lines). The two boundary redshifts are manually added to the sample. \textit{Bottom-left:} sampling without scattering of 10 metallicities for $z=2$. The product between the GSMF and SFMR yields the SFRD per logarithmic bin of stellar mass. This distribution is used to sample galactic stellar mass by quantiles, as with redshift. In the lack of scatter, the redshift-mass pair univocally determines the metallicity.}
    \label{fig:sfrd_grid_scatterless}
\end{figure*}

\begin{figure*}
    \centering
    \includegraphics[width=\textwidth]{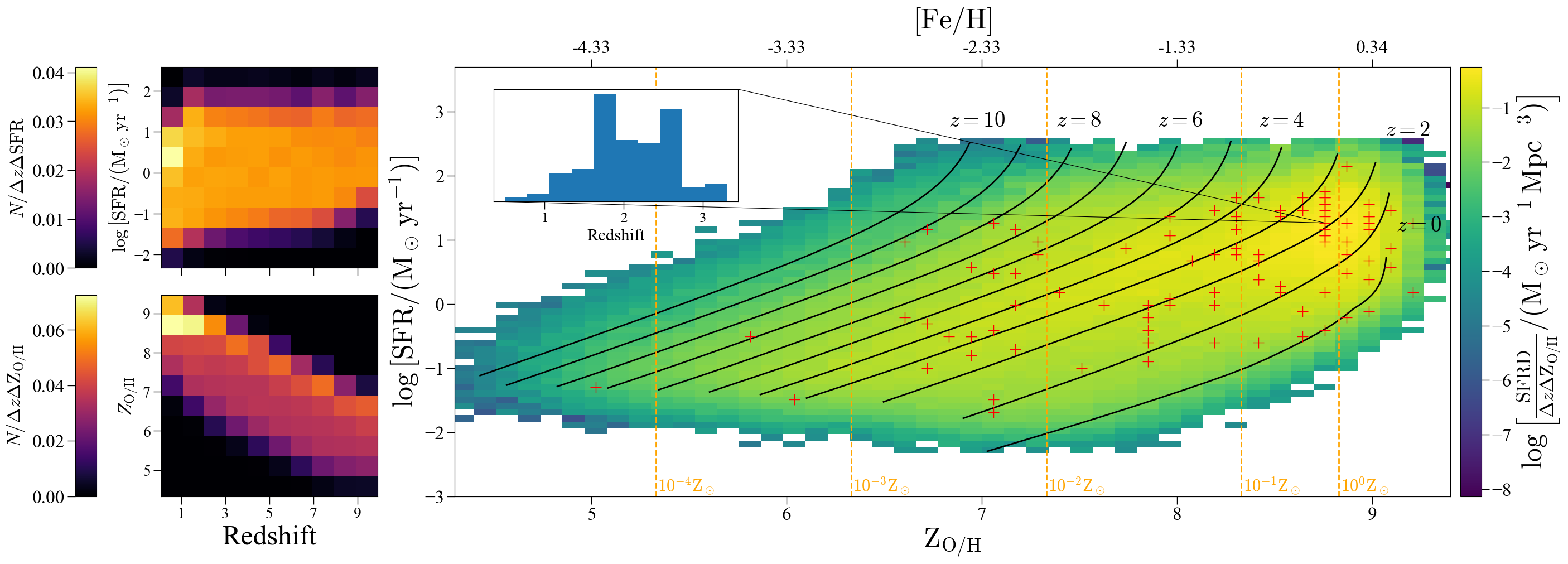}
    \caption{\textit{Right:} SFRD computed from the same model as in Fig. \ref{fig:sfrd_grid_scatterless}, but with the SFR and metallicity scatter model from \citet{Chruslinska2019}. The background SFRD was again generated from a sample of $250,000$ galaxies. The same lines of constant redshift (solid black lines) are retained for comparison, but here they only indicate the average properties of galaxies at different redshift. Dashed orange lines indicate metallicity in units of solar metallicity. The SFRD is computed within 2D bins of fixed widths $\Delta\log\SFR=0.095$ and $\Delta\ZOH=0.095$, and each bin corresponds to a particular redshift distribution due to scatter. The inset shows an example redshift distribution, for the peak SFRD bin. The red crosses show the result of sampling of 100 galaxies. Sampling is performed first on the SFR-metallicity plane with the SFRD as weight; then a redshift is drawn from the redshift distribution corresponding to the bin in which each galaxy falls. The full redshift distributions corresponding to the background plot are illustrated on the left. \textit{Top-left}: SFR-redshift distribution resulting from sampling with scatter. The shaded plot indicates the number of systems in a bin divided by its area, normalized so that all bins sum to one. \textit{Bottom-left}: the same as top-left, but for the metallicity-redshift distribution.}
    \label{fig:sfrd_grid_scatter}
\end{figure*}

For a population with known $\lrp{z,M_\ast}$, its metallicity ($\Z$) and SFR are given by the MZR and SFMR, respectively. For the MZR the greatest uncertainty is around the calibration of the conversion between observations and metallicity, and thus they collect four different calibrations: by \cite{Maiolino2008}, refined by \cite{Mannucci2009} (M09); and by \cite{Tremonti2004} (T04), \cite{Kobulnicky2004} (KK04) and \cite{Pettini2004} (PP04). Given evidence for a mass-dependent scatter, the metallicity is taken to be distributed normally about the MZR with $\sigma=0.1\,\mathrm{dex}$ for $M_\ast>10^{9.5}\,\Msun$, and linearly increasing with decreasing mass below that threshold. An additional normal scatter with $\sigma=0.14\,\mathrm{dex}$ is also included to account for the metallicity distribution within individual galaxies. For the SFMR, the main uncertainty is related to the degree to which the relation flattens at high $M_\ast$; therefore they build three models: Sharp, Moderate and No Flattening. All of the SFMR models are based on the model by \citet{Boogaard2018} in a lower mass region (varyingly defined), which the No Flattening model extends to the entire mass range. The Moderate and Sharp models base the higher mass region on the models by \citet{Speagle2014} and \citet{Tomczak2014}, respectively. Intrinsic scatter in the SFMR is also accounted for by assuming the SFR to be normally distributed.

The SFR measurements used to fit the SFMRs collected by \citetalias{Chruslinska2019} were taken under the assumption of a \citetalias{Kroupa2001} IMF when recovering the SFR from $\mathrm{H}\alpha$ observations; \citetalias{Chruslinska2020} use the \texttt{PÉGASE2} \citep{PEGASE1,PEGASE2} spectral synthesis code to compute corrections to the SFMRs for the IMF from \citetalias{Jerabkova2018}, which they make freely available. We thus we use the relations from \citetalias{Chruslinska2019} by themselves when working in the Invariant IMF model, and apply the corrections from \citetalias{Chruslinska2020} when working in the Varying IMF model. At this time we do not explicitly account for the fundamental metallicity relation (FMR), which expresses a tendency of galaxies of the same mass to have lower metallicities the greater their SFR is.

One last source of uncertainty is that, while the Varying IMF depends on the metallicity as $\FeHinline$, the MZR returns a metallicity $\ZOH=12+\log\lrp{\mathrm{O}/\mathrm{H}}$; it is thus necessary to convert between the two quantities, a process which is not at all well-defined. \citetalias{Chruslinska2020} propose two conversions: one which simply assumes [Fe/H]=[O/H]; and another which connects [Fe/H] and [O/H] through a two-part linear function, expressing in a simple way a greater initial abundance of oxygen, overtaken by iron on the timescale of type Ia supernovae occurrence. We use the second assumption as it is considered the more realistic between the two, although still not a precise description of chemical evolution \citep[but see][on the {[O/Fe]}-specific SFR of galaxies]{Chruslinska2024OFe}.

Between the GSMF, MZR and SFMR, there are a total of 24 possible permutations of the environmental conditions models. Of these, \citetalias{Chruslinska2019} defines the High and Low Metallicity extremes according to the overall shift of the cSFH with regard to metallicity: KK04 MZR, No Flattening SFMR and Fixed Slope GSMF (hereon "High Metallicity" model); and PP04 MZR, Sharp Flattening SFRM and Fixed Slope GSMF (hereon "Low Metallicity" model). At the approximate midpoint between the two they define a "moderate" permutation (hereon "Moderate Metallicity" model), with M09 MZR, Moderate Flattening SFMR and Fixed Slope GSMF. In relation to the Invariant model, the Varying cSFH has a longer low-metallicity tail at all redshifts, and sees an overall drop of the SFR at all but the lowest redshifts \citepalias[see][for a full discussion]{Chruslinska2020}. For the Invariant model, we compare the SFR density (SFRD) as a function of redshift obtained in the Moderate Metallicity model to that from \citet{SFHMadau} (who assume a \citetalias{Kroupa2001} IMF) in Fig. \ref{sec3fig:madau}. We find the models to be in good agreement, in particular in the "mean" case (without scatter), up to redshift $\sim8$, above which our model sharply drops due to a an abrupt decrease of the GSMF normalization in the \citet{Chruslinska2019} model (see their Figure 3). We point to Section 4 of \citet{Chruslinska2019} for a discussion of the SFRD from different model combinations and a comparison with previous well-established models, such as that due to \citet{SFHMadau}.

\section{Building the sampling pipeline}
\label{sec:3sampling}

While joining the distributions discussed in Sec. \ref{sec:2distributions} is a mostly straightforward process, there are some nuances which require more careful treatment, in particular with regard to the IMF; and also to how we treat the companion frequency and multiplicity, as discussed in Sec. \ref{sec2sub:multiplicity}. In the following we describe the structure of our initial sampling pipeline: in Sec. \ref{sec3sub:gal_sampling} we describe the sampling of "galaxies", i.e., environmental conditions for the simple populations that will make up the entire sample; in Sec. \ref{sec3sub:bin_sampling} we describe the sampling of individual systems within a simple population, with particular care for dealing with the IMF; and in \ref{sec3sub:pipeline_sampling} we summarize the pipeline.

\subsection{(Galaxy) environment sampling}
\label{sec3sub:gal_sampling}

\begin{figure*}
    \centering
    \includegraphics[width=\textwidth]{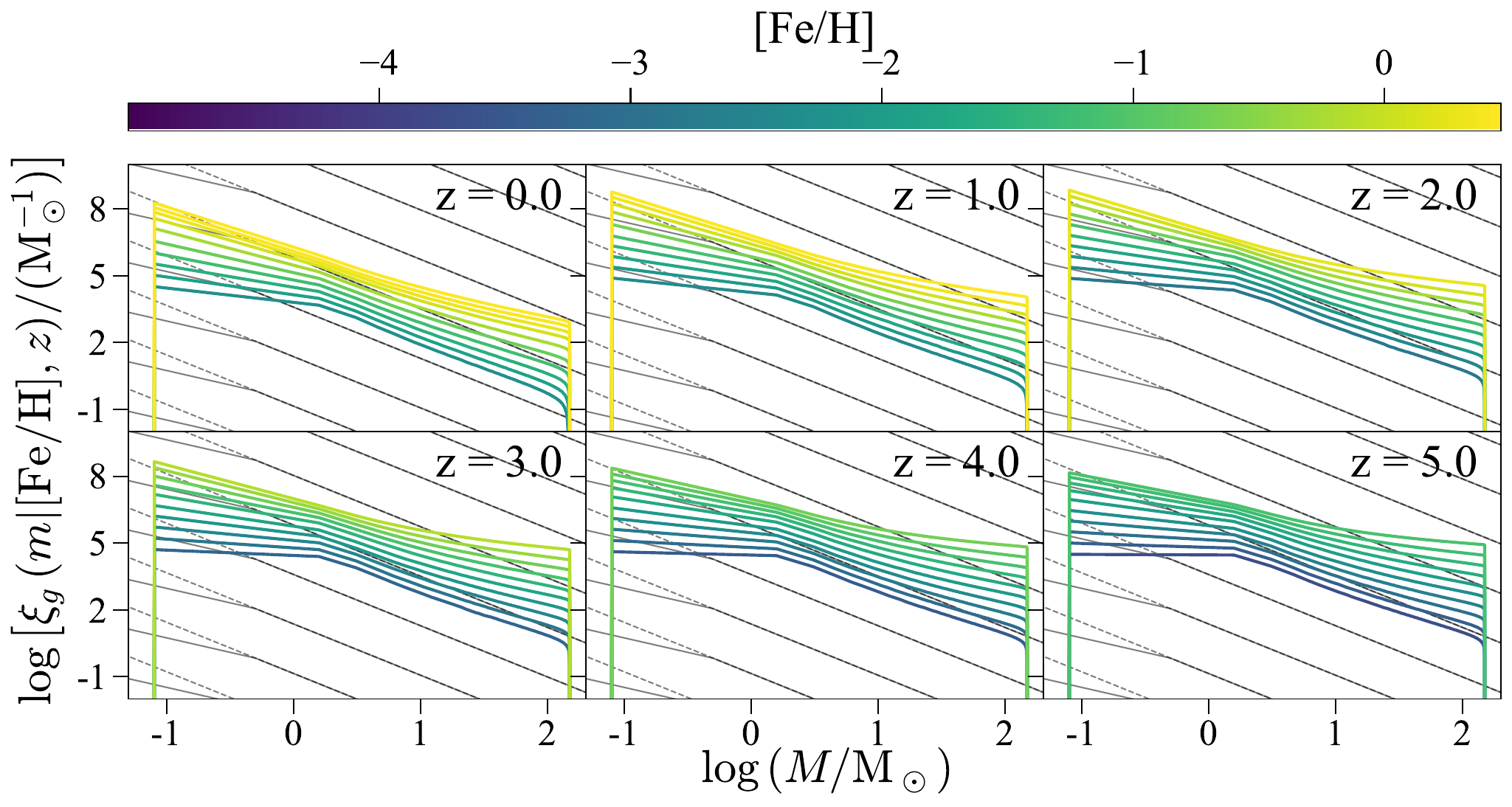}
    \caption{The gwIMF (equation \eqref{eq:gIMF_final}) computed following \citetalias{Jerabkova2018} (colored lines) for different redshift and metallicity combinations, here treated as independent variables. The \citet{Salpeter1955} (dashed black lines) and \citetalias{Kroupa2001} (solid black lines) IMFs are shown for comparison. We see the IMF become strongly top-heavy with increasing redshift, and bottom light with decreasing metallicity. However, it appears to become top-heavy with high metallicity. This is a consequence of high metallicities being linked to high star-formation rates, which make the IMF top-heavy --- see Figs. \ref{fig:sfrd_grid_scatter} and \ref{fig:sfr_imfs}. This confusion could perhaps be mitigated by implementing the FMR --- see Sec. \ref{sec3sub:imf_sampling}  Each curve is normalized to $\SFR\times10\,\mathrm{Myr}$, for the SFR correspondent to each $\lrp{z,\Z}$.}
    \label{fig:redshift_imfs}
\end{figure*}

As we intend for each initial sample to be representative of the entire span of cosmic star formation history under a particular model, a sample is always a \textit{composite} population, wherein each \textit{simple} stellar population (SSP) is generated for a given $\lrp{z,\Z,\SFR}$ (a "galaxy"), connected through the distributions in Sec. \ref{sec2sub:environment}. As the IMF depends only on two parameters, we will often identify the initial conditions of each SSP as $\lrp{z,\Z}$.

We implement two different sampling methods in \texttt{BOSSA}, depending on whether scatter is accounted for or not in the SFMR and MZR; both methods are weighted by the star-formation rate density (SFRD) by default, but options for weighting by galaxy mass or number density (computed from the GSMF) are also available in the scatterless case. The two pipelines are best summarized by Figs. \ref{fig:sfrd_grid_scatterless} (scatterless sampling) and \ref{fig:sfrd_grid_scatter} (sampling with scatter).

In scatterless sampling, the SFRD, and the SFRD per logarithmic galaxy stellar mass bin, is calculated as

\begin{equation}
    \label{eq:sfrd_def}
    \SFRD(z) = \int \overbrace{\underbrace{\frac{\d n_\mathrm{gal}}{\d V_\mathrm{c} \d\log M_\ast}}_{\mathrm{GSMF}} \underbrace{\frac{\d m}{\d t}}_{\mathrm{SFMR}}}^{\SFRD/\Delta\log M_\ast}\, \d\log M_\ast.
\end{equation}

The sampling proceeds as follows,

\begin{enumerate}
    \item A set of $n_z$ redshift values is drawn from the integrated $\SFRD(z)$ by finding its $n_z$ quantiles (redshift bins which all contain the same total SFRD), and representing each one by its SFRD-weighted average redshift. This is illustrated in the top-left of Fig. \ref{fig:sfrd_grid_scatterless}.
    \item For each redshift, sample $n_\Z$ metallicities (determining the Varying IMF), by first drawing $n_\Z$ galaxy stellar masses from the SFRD per logarithmic mass bin through the quantile method (bottom-left of Fig. \ref{fig:sfrd_grid_scatterless}). Then, assign each mass a metallicity, as well as SFR, from the MZR and SFMR.
\end{enumerate}

\noindent With analogous steps for GSMF-based weighting.

The right panel of Fig. \ref{fig:sfrd_grid_scatterless} shows the resulting sample for $n_z=10$ and $n_\Z=10$, within the $z=0-10$ and $\log M_\ast/\Msun=6-12$ ranges, and with the manual inclusion of $z=0.01$ and $10$ to the sample, under the Moderate Metallicity and Varying IMF models. The inclusion of $z=0.01,10$ is also meant to yield boundary conditions for this approximation. The main feature of a scatterless approach is the univocal relation between redshift and SFR, metallicity pairs, which can greatly facilitate the computation of quantities that vary over redshift, such as merger rates. However, we caution that this approach should only be employed when absolutely necessary, or for cases where evaluating the effect of model variations is the focus, rather than the accuracy of the final results. Otherwise, the assumption of a univocal redshift-SFR,metallicity relationship may lead to erroneous inferences from observations about how merger properties evolve with redshift, and it ignores work that has and continues to be done on constraining the scatter in galaxy empirical relations \citep[e.g.,][]{scatterBerti2021,scatterSherman2021,scatterLi2023,scatterGarcia2024,scatterPistis2024}.

We therefore also implemented a sampling with scatter approach as a more generally applicable and physically accurate option, which requires a different approach due to the loss of one-to-one correspondence between a metallicity-SFR pair and redshift. In this case, we proceed similarly to \citet{Chruslinska2019}, as follows,

\begin{enumerate}
    \item The redshift range is divided in $n_z$ uniform bins. Then, the mass range is divided into $n_M$ uniform bins. For each redshift bin, the total galaxy density $n_\mathrm{gal}$ within each mass bin is calculated by integrating the GSMF. The GSMF is assumed to be constant within each redshift bin.
    \item For each mass-redshift bin, a sample of $N_\mathrm{gal}$ masses are drawn, each one representing a total galaxy density $n_\mathrm{gal}/N_\mathrm{gal}$. Each galaxy is assigned a metallicity and SFR, drawn from the MZR and SFMR under the assumption of normal scatter.
    \item All galaxies are then collected into metallicity-SFR bins, and the total SFRD within is calculated by summing their individual contributions, $\SFR\times n_\mathrm{gal}/N_\mathrm{gal}$. The set of redshifts of the galaxies falling within each bin then defines that bin's redshift distribution.
    \item A physical sample is generated by drawing galaxies from the SFR-metallicity bins with the SFRD as weight, and assigning them a redshift drawn from the redshift distribution within their respective bin.
\end{enumerate}

The right panel of Fig. \ref{fig:sfrd_grid_scatter} illustrates a sample of 100 galaxies (crosses) resulting from this method, over the SFRD distribution computed with scatter, under the same model assumptions and the same redshift and mass ranges as with Fig. \ref{fig:sfrd_grid_scatterless}. The left panels of Fig. \ref{fig:sfrd_grid_scatter} show also the corresponding redshift-SFR (top) and redshift-metallicity (bottom) distributions. The impact of scatter is clear, and highlights the shortcomings of the scatterless approach, especially with regard to the more complicate, physical relation between redshift, metallicity and SFR.

\subsection{Mass and orbital parameters sampling}
\label{sec3sub:bin_sampling}

\subsubsection{How to sample the IMF?}
\label{sec3sub:optimal_sampling}

So far our treatment of the IMF as a probability density function (PDF) of stellar masses for a given star formation event has adhered strictly to the definition provided in Sec. \ref{sec2sub:imf}. However, although in practical terms this is the most adequate interpretation of the IMF for our purposes, it is important, since we are working within the IGIMF framework, to mention and briefly discuss how this traditional interpretation may not be in fact the most accurate to the physical process of star formation, or when describing it from the IGIMF approach.

As discussed in depth by \citet{Kroupa2013} and \citep{Kroupa_Jerabkova_2021}, treating the IMF as a PDF establishes a purely probabilistic view of star formation, as a stochastic process where even two identical ECLs will in general produce different star populations. While the PDF treatment is commonplace, the modern picture of star formation relies to at least some level on the presence of self-regulated mechanisms to explain its characteristic timescales, which are much longer than simply the free fall times of molecular cloud cores; supersonic turbulence, for example, is believed to both help support molecular clouds against collapse as well as seed localized cores along shock fronts. This implies on a process that is not purely stochastic, and when brought to an extreme points to a \textit{deterministic} view of star formation.

\textit{Optimal sampling} is a sampling method developed within IGIMF theory in order to represent the fully deterministic view, such that the population resulting from a given gwIMF is well-defined (an \textit{optimal} sample), while still respecting the IGIMF constraints (equations \eqref{eq:imfconstraint1_totmass} and \eqref{eq:imfconstraint2_mmax}). Comparison with observations have favored the optimal sampling of the IGIMF over pure random sampling \citep[e.g.,][]{Weidner2013_mmax,Weidner2014}, although this does not necessarily imply on the IMF being what has been termed an optimal distribution function (ODF); rather, it might be the case that it stands somewhere between the pure PDF and ODF extremes. The role of stochasticity and self-regulation in star formation thus remains an open question \citep[see, e.g.,][]{Eldridge2012stoch,Stanway2023stoch,Yan2023stoch,Grudic2023stoch}.

We apply here, at any rate, pure random sampling to the IMF as a PDF, unless stated otherwise. Although this may not be adequate for the precise generation of particular population, it suffices at least as first approximation for capturing average population properties when our focus is their evolution over time \citep[although][conclude star formation cannot be a random process, they agree that it can be a good first approximation in many cases]{Kroupa2013}.

\subsubsection{What to sample from the IMF?}
\label{sec3sub:imf_sampling}

The sampling of masses calls for particular care, since there are always at least two sets of masses involved: that of primaries masses, $\lrc{m_1}$; and that of their companion masses, $\lrc{m_i}$. While it is an established practice in binary population synthesis to sample $\lrc{m_1}$ from the IMF, and then select its companion masses from the mass ratio distribution, this results in only $\lrc{m_1}$ reproducing the IMF, which is not necessarily consistent with how the IMF is defined or measured. We propose here that a more critical discussion on the exact meaning of the IMF is relevant for binary population synthesis. 

There are generally two ways of defining the IMF, which are not always incompatible, but do require clear delineation of what kind of systems the definition is concerned with. In one, which we might term an \textit{a priori} definition, the IMF is taken as a physical distribution emerging from some universal process or set of processes that set its shape across all star formation events. Turbulence has often been pointed to as a dominant, or at least significant, part of such processes \citep[see, e.g.][]{Padoan2002turbIMF,Hennebelle2008turbIMF,Chabrier2011turbIMF,Haugbolle2018turbIMF,Mathew2023turbIMF}, although other mechanisms have argued for \citep[e.g.,][]{Andre2014turbIMF,BertelliMotta2016turbIMF,Liptai2017turbIMF}. The possibility that turbulence has a dominant rule in setting the IMF is particularly attractive for the question of IMF variability, as it would fit as a \textit{universal} mechanism capable of setting approximately the same IMF for all star formation events (see, e.g., \citeauthor{Hopkins2013univIMF}, \citeyear{Hopkins2013univIMF}; and the review by \citeauthor{Offner2014univIMF}, \citeyear{Offner2014univIMF}).
 
Alternatively, in a \textit{a posteriori} definition, the IMF may be defined in terms of how it is measured, i.e., it describes the initial masses of those populations for which it has been constrained. If this definition is to be taken rigorously, then we are severely limited. From first principles, the IMF may not even be considered as an observable itself: if it is a distribution that should arise from a single stellar-formation event, than it never fully does, as the most massive components of a stellar population will have already evolved off of the main sequence by the time the least massive components have reached ZAMS, in addition to the effects of binary evolution and stellar feedback on the shape of the population \citep[see, e.g.,][]{deMink2014,Schneider2015,Oh2018}. This is discussed as the "IMF Unmeasurability Theorem" in \citet{Kroupa2013}. Measuring (or deducing) the IMF must then make simultaneous use of stellar populations of different ages, all while taking careful account of any effects that might already have deviated their present-day mass function from the IMF. In terms of observations as well, low-mass and massive stars pose entirely different challenges in terms of population sizes, detectability and, key for our case, multiplicity, as stars become increasingly dominated by multiple systems with increasing mass (Sec. \ref{sec2sub:multiplicity}). This issue was considered by \citetalias{Kroupa2001} in reanalyzing the extensive sample of populations for which the IMF slope had been measured from \citet{Scalo1986} and considering possible sources of non-physical (i.e., due to observational effects) variations of the IMF previously unaccounted for, including unresolved binaries. This and later work has found that the IMF is significantly affected by this bias only for systems with $\lesssim1\,\Msun$ in total mass \citep{Kroupa_Jerabkova_2021}. In terms of \textit{component} masses, it has been argued that the IMF should be understood as corresponding to \textit{all} stars formed from a single star formation event, both isolated, primaries and companions \citep{Kroupa1995binimf,Kroupa2013}. Due to the role of pre-ZAMS evolution, however, this does not map clearly to the ZAMS populations used as starting conditions in BPS; in light of this, some works make an important distinction between \textit{birth} and \textit{initial} (ZAMS) populations, only the former of which would be directly set by the IMF.

This latter picture (which we will refer to as "empirical") clearly defines the appropriate sampling of a \textit{birth} binary population, wherein all component masses are sampled from the IMF, but paired according to a mass ratio distribution, a procedure which has been advanced in \citet{oh2012,oh2016} and \citep{oh2015}. On the other hand, within the former picture (which we will refer to as the "universal mechanism" picture), as discussed in Sec. \ref{sec2sub:period}, the formation of multiples can be expected to occur not only through filament/core fragmentation, but also through disk fragmentation. In this latter case, dynamical interactions aside, the IMF clearly applies to populations of isolated stars, where there is no further fragmentation; but if turbulence, or another combination of universal mechanisms, initially sets up the spectrum of prestellar core masses, which then further fragment and have their fragments co-evolve before ZAMS, we have no reason to expect the final ZAMS mass distribution to retain its pre-stellar core-set shape. In this instance it is clear that the IMF applies to the mass of \textit{primary} stars, and perhaps only if they are isolated.

We do not propose ourselves here to tackle this question of the very physical content of the IMF, as already has and continues to be done by some of the aforementioned works here and in Sec. \ref{sec2sub:imf}. Instead, we merely point to the potential inconsistency in sampling $\lrc{m_1}$ from the IMF, and the mass of its companion(s) from $\mathcal{P}_\mathrm{q}$ independently. We thus propose a method to retain \textit{consistency} in the sampling, without making claims about the process of star formation.

In the "universal mechanism" picture, one method might be to sample \textit{total masses} from the IMF, and break them into component masses from $\mathcal{P}_\mathrm{q}$. Here, however, we choose the more conservative approach of the "empirical" picture: the full set of primary masses $m_1$ and companion masses $m_i$ must reproduce the IMF, and be paired according to $\mathcal{P}_\mathrm{q}$. This is analogous to the method in \citet{oh2016} but, of course, ignores how the mass function is affected by pre-ZAMS evolution, which we cannot model at this point \citep[but see][]{belloni2017}. For further discussion on how the stellar IMF is impacted by binaries, we refer the reader to \citet{kroupajerabkova2018impact}.

We start by sampling a single \textit{mass pool} from the IMF, from which we will draw all further masses. Then, given a chosen $\ncp^\mathrm{max}$ and a mass tolerance $\delta m$, the steps for generating a binary are as follows,

\begin{enumerate}
    \item Randomly draw a $m_1$ from the mass pool,
    \item Draw a $\ncp$ from the chosen $\mathcal{P}_{\ncp}\lrp{m_1}$,
    \item For each of the $\ncp$ companions,
    \begin{enumerate}
        \item Draw a $\log P$ from $\mathcal{P}_{\log P}$,
        \item Draw $q$ and $e$ from $\mathcal{P}_q$ and $\mathcal{P}_e$. This defines the drawn companion mass, $m_\mathrm{cp}^\mathrm{drawn}=qm_1$
        \item Find the closest $m_\mathrm{cp}^\mathrm{pool}$ to $m_\mathrm{cp}^\mathrm{drawn}$ in the mass pool.
        \item If $\abs{m_\mathrm{cp}^\mathrm{pool}-m_\mathrm{cp}^\mathrm{drawn}}/m_\mathrm{cp}^\mathrm{pool}\leq\delta m$, store the companion and continue. Otherwise, return to step 1.
    \end{enumerate}
    \item With all companions successfully sampled, store the system and remove all sampled masses from the sampling pool.
\end{enumerate}

This ensures sampling of masses without repetition from a list on which the shape of the IMF is imposed, such that we expect to preserve the IMF while pairing masses according to the mass ratio distribution. These steps apply to both the Invariant and Varying orbital parameter distributions. In Fig. \ref{fig:redshift_imfs} we show the Varying IMF for different metallicity-redshift pairs; it generally becomes bottom-light with decreasing metallicity, and top-heavy with increasing redshift \textit{and metallicity}. This is a consequence of high average metallicities being, in general, associated to high average SFRs (Figs. \ref{fig:sfrd_grid_scatter} and \ref{fig:sfrd_grid_scatterless}), while increasing SFRs make the IMF top-heavy much more intensely than decreasing metallicities make it top-heavy (Fig. \ref{fig:sfr_imfs}). This is gives the appearance of a behavior opposite to what is expected, something which could perhaps be mitigated by including the FMR.

\subsection{The sampling pipeline}
\label{sec3sub:pipeline_sampling}

\begin{table}
	\centering
	\caption{Model options included in our pipeline.}
	\label{tab:models}
	\begin{tabular}{lr} % four columns, alignment for each
		\hline
		Name & Options \\
		\hline
            \multicolumn{2}{c}{Environmental conditions} \\
		GSMF & Fixed or varying low-mass slope \\
		MZR & T04, M09, KK04 or PP04 \\
            SFMR & No , moderate or strong flattening \\
            \multicolumn{2}{c}{Binary parameters} \\
            IMF & Invariant or Varying  \\
            Orbital parameters & Invariant or Varying distributions  \\
            Multiplicity & All Multiples (AM) or Only Binaries (OB) \\
		\hline
	\end{tabular}
\end{table}

Our full set of initial condition models is composed of 192 possible permutations, summarized in Tab. \ref{tab:models}. When sampling $\ncp$ for a primary of mass $m_1$, the probability of each $\ncp$ is given by its respective multiplicity fraction at that mass, according to either the OB or AM model. When bulding the mass pool, we draw masses from the full $[0.08\,\Msun,150\,\Msun]$ range; although we never draw primaries $m_1<0.8\,\Msun$, this allows asymmetric pairs with $m_1<8\,\Msun$ to be drawn. In summary, the full initial sampling pipeline is as follows,

\begin{enumerate}
    \item Sample $N$ SSPs, each labeled by its $\lrp{z,\Z,\SFR}$, through either the scatterless or with scatter methods (Sec. \ref{sec3sub:gal_sampling}).
    \item For each simple SSP,
    \begin{enumerate}
        \item Draw a mass pool of size $N_m$ from the IMF,
        \item Sample multiples up to order $\ncp$ until either,
        \begin{enumerate}
            \item The mass pool is exhausted, or
            \item $N_\mathrm{max}$ failed iterations pass.
        \end{enumerate}
    \end{enumerate}
\end{enumerate}

A further point of variation would be with regard to the $\ZOH\to\FeHinline$ conversion; for now, we always implement the conversion which takes into account in a simple way the different timescales for oxygen and iron enrichment (Sec. \ref{sec2sub:environment}).

\subsubsection{Completeness of the sample}
\label{sec3sub:completeness}

In order to adequately normalize our binary samples, we need to carefully account for both our assumptions and for the validity ranges of the distributions we implement. For galaxy stellar masses, we sample between $10^6$ and $10^{12}\,\Msun$ and impose no limits on metallicity and SFR besides those that naturally emerge from that range through the MZR and SFMR. We consider star formation outside of this range to be negligible. For star masses, we consider that stars form between $0.08$ and $150\,\Msun$, but we sample only above $0.8\,\Msun$ \citep{MoeDistefano2017}. For companions, we only sample mass ratios between $0.1$ and $1.0$; thus, we miss also some companion stars in the range $0.08-15\,\Msun$, and all $m_1<0.8\,\Msun$ primaries.

Our mass sampling method (Sec. \ref{sec3sub:imf_sampling}) incurs extra "losses" because the mass pool is never in fact exhausted, and thus the final mass sample never corresponds \textit{exactly} to a sample randomly drawn from the IMF. As masses are removed from the sampling pool, the number of valid companion masses (those that pass the tolerance test in Sec. \ref{sec3sub:imf_sampling}) for any $m_1$ becomes increasingly smaller; eventually none remain, and the sampling stops. Because massive stars are less numerous and $q\geq0.1$, the leftover masses are mostly close to the upper end of the sample. 

In practice, we can consider these limitations as negligible as long as the full sample of masses continues to recover the original shape of the IMF in the $\lrs{0.8\,\Msun,150\,\Msun}$. In this case we can estimate the total $M_\mathrm{sf}$ of the population by appropriately normalizing the IMF to that range. We verify this in Sec. \ref{sec:4results}.

\section{Consistency checks and emergent behavior}
\label{sec:4results}

Below we examine initial samples generated with \texttt{BOSSA} for a few different model settings in order to verify our sampling pipeline and check for its implications for the mass and orbital parameter distributions. For every model combination cited, we draw $10$ redshifts and $10$ metallicities; and for each simple population we start from a pool of $4\times10^6$ masses, which results in $\sim10^6$ systems being drawn. Thus each model contains $\sim10^8$ systems in total. Environmental condition sampling is always done from the Moderate Metallicity distributions.

\subsection{Mass distributions}
\label{sec4sub:mass_distr}

\begin{figure}
    \centering
    \includegraphics[width=\columnwidth]{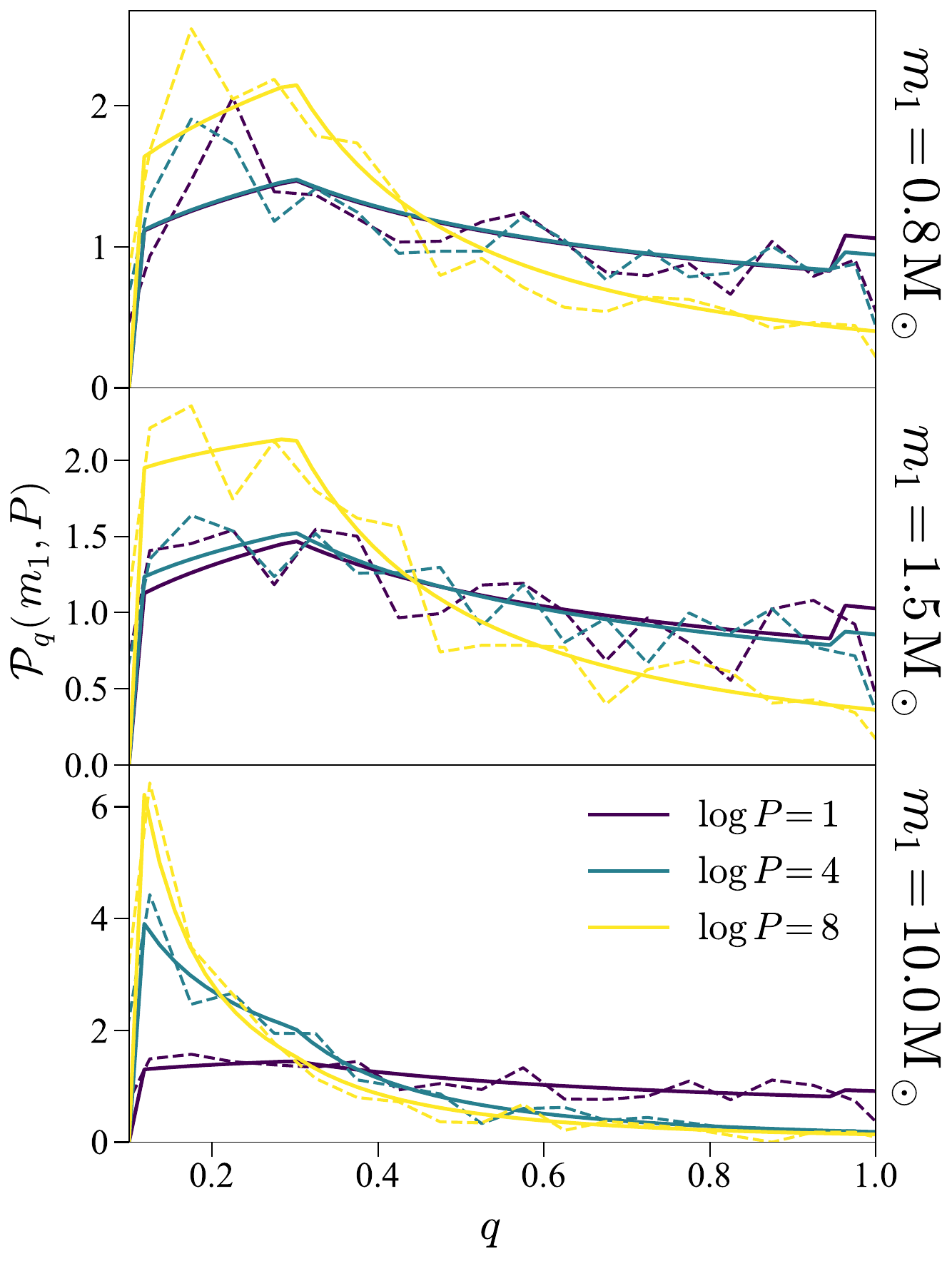}
    \caption{Initial mass ratio distributions for a small set of $m_1$ and $\log P$ from the original analytical form by \citet{MoeDistefano2017} (solid lines) and its discrete form resulting from the sampling procedure described in Sec. \ref{sec3sub:imf_sampling} (dashed lines). Employing the $q$ distribution as a mass-pairing rule instead of directly sampling $q$ preserves the distribution shape within the sample resolution. Here we used a sample with our standard settings: 1,000 binaries taking one of 100 possible values.}
    \label{fig:q_sampling_check}
\end{figure}

We first verify that the shapes of the IMF and of $\mathcal{P}_q$ are preserved in the sample. In Fig. \ref{fig:q_sampling_check}, we compare the analytical Varying distributions to that resulting from \texttt{BOSSA} for a few different cases; the plot presents a sub-sample of systems within $5\%$ of each $\lrp{m_1,\logP}$  pair, which limits their resolution. We can verify that the mass ratio distribution is preserved, including in the close-to-uniform case, similar to the Invariant distribution.

\begin{figure*}
    \centering
    \includegraphics[width=\textwidth]{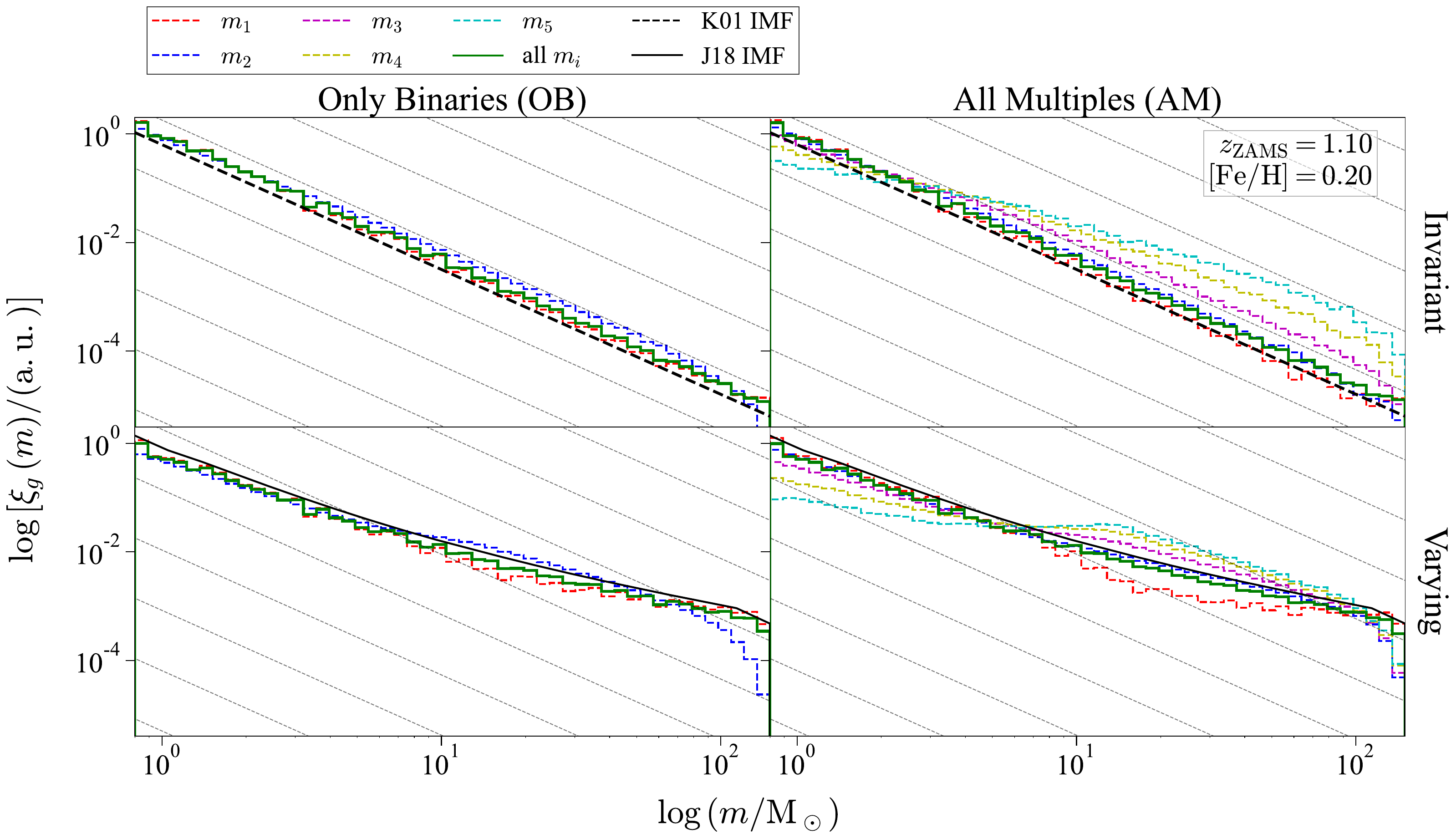}
    \caption{Result of our mass and mass ratio sampling procedure (Sec. \ref{sec3sub:imf_sampling}), in a simple population example ($z=1.1,\FeHinline=0.2$), for four models. Clockwise, starting from the top-left, these are: Invariant+OB, Invariant+AM, Varying+AM and Varying+OB. We show the IMF resulting for individual components (colored dashed lines) as well as the full IMF resulting from the sampling (green solid line), and the original IMF from which the sample was drawn, either the \citetalias{Jerabkova2018} (solid black line) or the \citetalias{Kroupa2001} (thick dashed black line) IMF. The \citetalias{Kroupa2001} IMF is always shown for reference (thin dashed black lines). Companion indices reflect their orbital period, not mass ($m_2$ is always the mass of the closest companion and $m_5$ of the farthest). Note that the IMF for $m_i$ includes all systems with up to at least $i$ number of components. While the full sampled IMF reproduces the original IMF in all cases, the individual component IMFs do not, particularly in the AM cases. In the Invariant cases, the companion distributions are generally shifted towards larger masses because multiplicity tends to grow with $m_1$, and the mass ratio distribution is invariant. In the Varying cases, this effect competes with the tendency of further away companions being less massive (see Sec. \ref{sec2sub:massratio} for further discussion). }
    \label{fig:sampled_imfs}
\end{figure*}

In the following we do the same for the IMF, which holds some interesting new behavior. For an example $\lrp{z,\Z}$, we see in Fig. \ref{fig:sampled_imfs} the resulting samples for the four combinations of Invariant (top panels) and Varying (bottom panels), and OB and AM models, in comparison to the original IMFs; the IMF for each mass $m_i$ is generated from all masses of components of order $i$, i.e., the primary IMF is computed from all primary masses (in isolated stars, binaries, triples...), while the tertiary IMF is computed from all triple or higher-order systems. In each case, we can see that the full set of sampled masses (\textit{all $m_i$}) always reproduces the IMF, while the individual sets of primary, secondary, tertiary... masses do not; the higher the multiplicity, the more they deviate.

\begin{figure*}
    \centering
    \includegraphics[width=\textwidth]{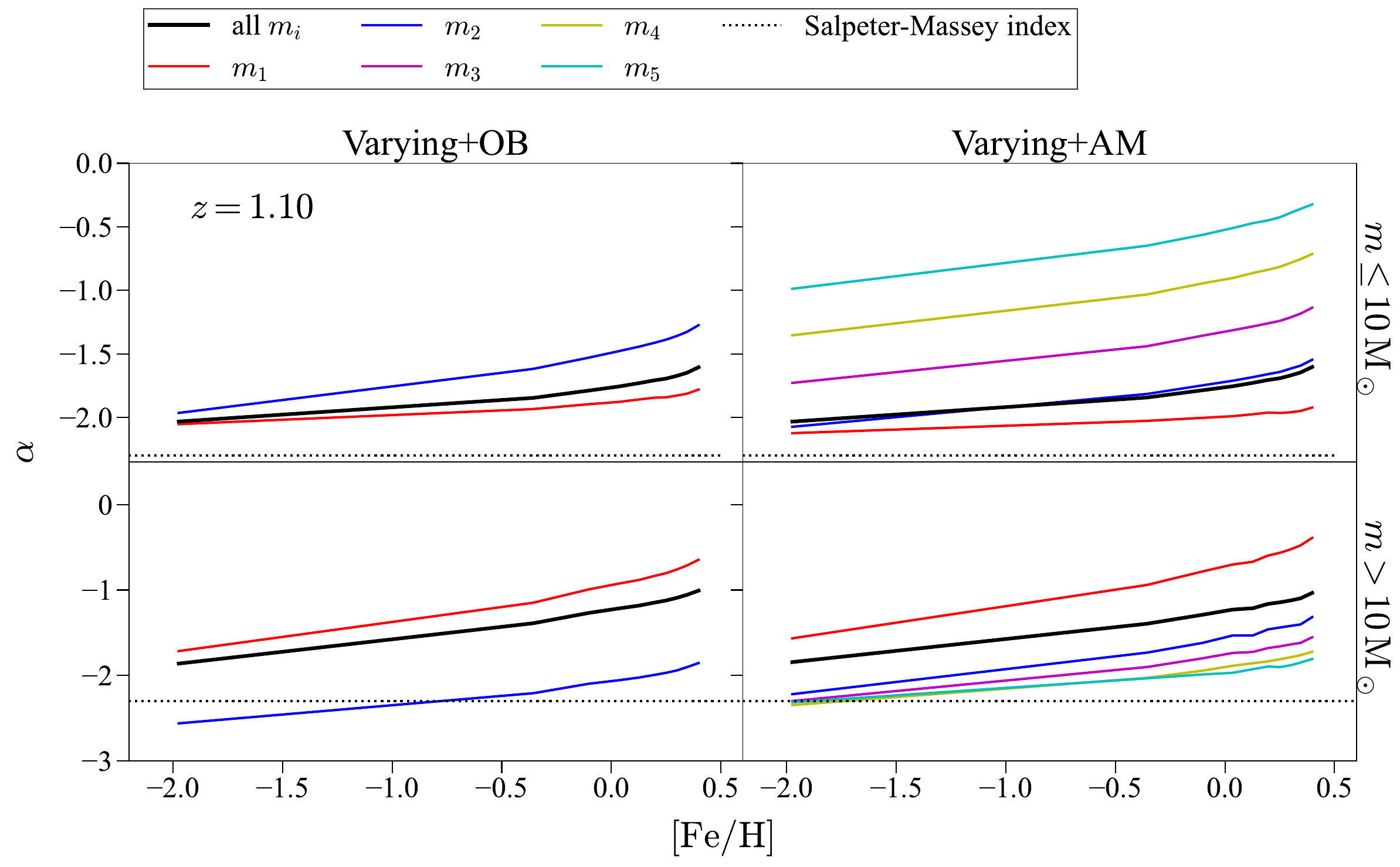}
    \caption{Best-fit power law slope ($\alpha$) for $m\leq10\,\Msun$ (top row) and $m>10\,\Msun$ (bottom row), for both the Varying+All Multiples (AM, left column) and Varying+Only Binaries (OB, right column) model combinations, for all metallicities sampled for the example $z=1.10$ redshift. We fit the slope for each component mass ($m_i$; for companions, greater $i$ indicates greater separation from the primary) distribution individually (colored curves), and also for the \citetalias{Jerabkova2018} IMF (thick solid black line) at each metallicity; we also show the Salpeter-Massey index $\alpha=-2.3$ for reference (thin dotted black line). All cases become overall flatter with increasing redshift, which indicates an increasingly bottom-light (flatter for $\leq10\,\Msun$) and top-heavy (flatter for $>10\,\Msun$) IMF; as discussed in Sec. \ref{sec3sub:imf_sampling}, this is a consequence of higher metallicities being associated with higher SFRs on average. In both the AM and OB cases, the primary IMF is steeper than the full IMF below $10\,\Msun$ and flatter above; the companion IMFs show the inverse behavior, which is the more pronounced the farther the companion is.}
    \label{fig:imf_onlybin_exponents}
\end{figure*}

There are two effects at play in setting the shape of the IMFs. The first is the tendency of $\ncp$ to increase with $m_1$: the IMFs of higher-order companions are automatically shifted towards higher masses in relation to the primary because the primary IMF corresponds to all primaries, while higher-order companions correspond only to more massive primaries, for which the companions themselves tend to be more massive ($q\geq0.1$). The second is the correlation between $P$ and $q$ for Varying orbital parameters, in which, as seen in Fig. \ref{fig:md17_qdistr}, longer orbital periods always shift mass ratios towards $0.1$. Thus, for samples generated in the Invariant model, the IMFs tend to become flatter overall for higher-order companions: they are more likely to be massive then a randomly selected primary (from systems of any multiplicity; within a given system the primary is always the most massive). For samples generated in the Varying model, on the other hand, the very massive companions are strongly suppressed, leading to a flatter pattern then the full IMF for $\lesssim10\,\Msun$, and steeper for $\gtrsim10\,\Msun$; when looking at the overall population, primaries are more likely to be below $\sim10\,\Msun$, and the higher-order a companion is, the more massive it tends to be; however, when looking only at $>10\,\Msun$ stars, primaries are on average more massive. This behavior can be seen more clearly in Fig. \ref{fig:imf_onlybin_exponents}, which shows the indices resulting from a power-law fit to the IMF in the regions below and above $10\,\Msun$, as a function of metallicity ,in all cases from Fig. \ref{fig:sampled_imfs}. There we see see the overbearing influence of increasing SFRs for higher metallicity, making the IMF flatter. For other redshift, the full IMF varies as in Fig. \ref{fig:redshift_imfs}, and the way in which it is broken down into the component IMFs does not change significantly.

With regard to the metallicity-independent Invariant model, we report the $\leq10\,\Msun$ and $>10\,\Msun$ best-fit slopes for the component IMFs in both AM and OB models in Table \ref{tab:imf_slopes}, which follow from the cases shown in Fig. \ref{fig:sampled_imfs}.

\begin{table}
	\centering
	\caption{Best-fit power law slopes ($\alpha$) for each component mass ($m_i$) in the $\leq10\,\Msun$ and $>10\,\Msun$ ranges, for the Invariant model, in both the OB and AM models.}
	\label{tab:imf_slopes}
	\begin{tabular}{lcccc} % four columns, alignment for each
		\hline
        Component & \multicolumn{2}{c}{All Multiples} & \multicolumn{2}{c}{Only Binaries} \\
		{} & $\alpha_{\leq10\,\Msun}$  & $\alpha_{>10\,\Msun}$ & $\alpha_{\leq10\,\Msun}$  & $\alpha_{>10\,\Msun}$ \\
		\hline
		$m_1$ & -2.44 & -2.34 & -2.36 & -2.28 \\
		$m_2$ & -2.18 & -2.50 & -2.08 & -2.55 \\
        $m_3$ & -1.80 & -2.38 & - & - \\
        $m_4$ & -1.41 & -2.14 & - & - \\
        $m_5$ & -1.08 & -1.90 & - & - \\
		\hline
	\end{tabular}
\end{table}

Finally, we confirm that, in spite of the limitations of our mass sampling discussed in Sec. \ref{sec3sub:completeness}, the final complete mass sample is in excellent agreement with the original IMF, as seen in all cases at Fig. \ref{fig:sampled_imfs}. Although we do not show the result for all $\lrp{z,Z}$ here, we have checked that this is always the case.

\subsection{Multiplicity fractions}
\label{sec4sub:multiplicity}

\begin{figure}
    \centering
    \includegraphics[width=\columnwidth]{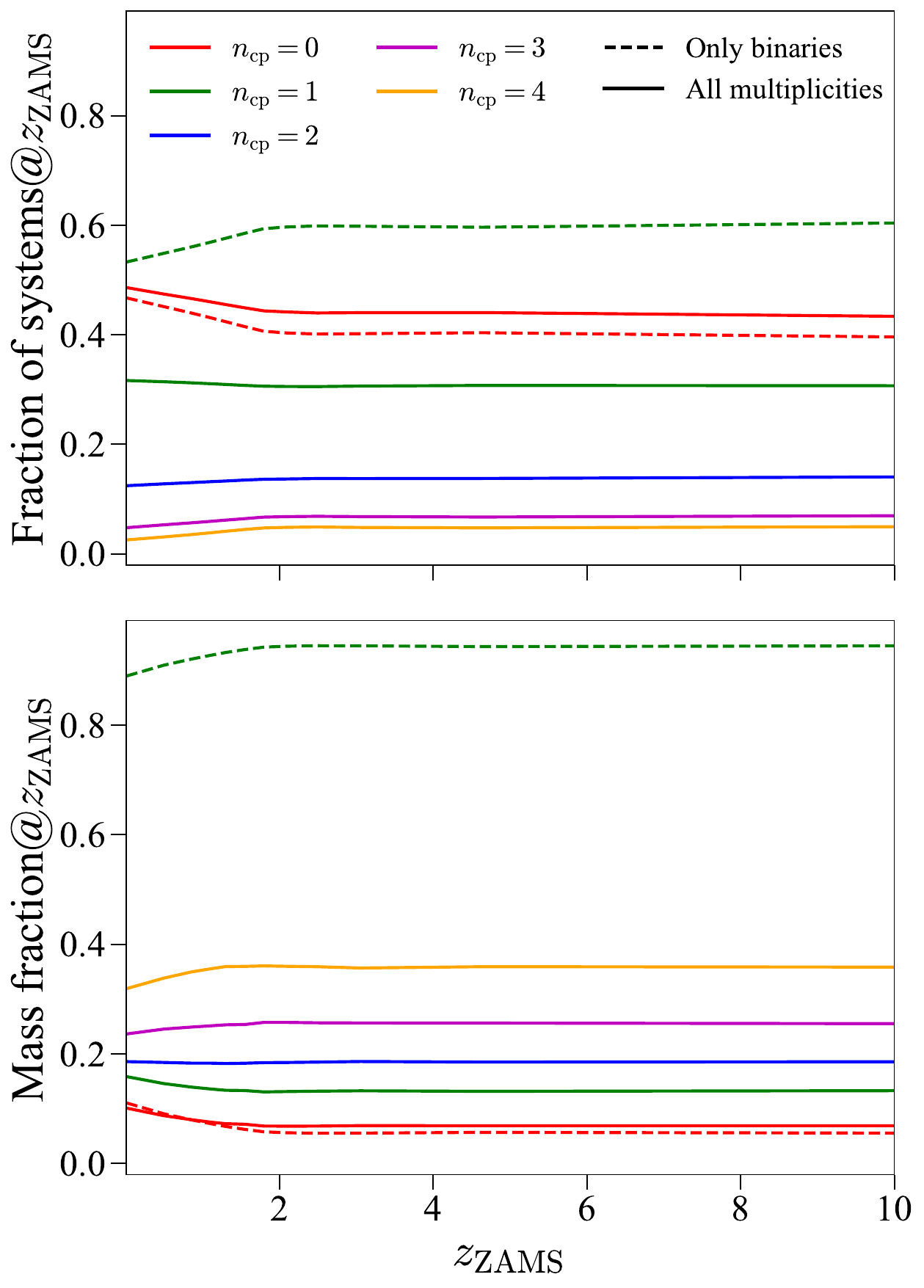}
    \caption{Number (top panel) and mass (bottom panel) fractions of systems with $m_1=0.8-150\,\Msun$ and $\ncp=0-4$ over redshift, for the Varying model, for both the OB and AM models (dashed and solid lines, respectively). Significant variation only happens below $z=2$, where the production of solar-type stars is enhanced in relation to that of massive stars, increasing the contribution of isolated primaries. The lower the multiplicity, the greater the number fraction, but the lower the mass fraction. Moving from the OB model impacts most strongly the binary mass fraction.}
    \label{fig:number_mass_frac}
\end{figure}

Although analytical orbital parameter distributions are not yet at the point where they describe any cosmic evolution, it is interesting to check whether the coupling of the Varying orbital parameter distributions to the Varying IMF induces any such dependency. In Fig. \ref{fig:number_mass_frac} we verify that this is not the case with regard to redshift for either the number fraction (top panel) or mass fraction (bottom panel), considering all sampled primaries, from $m_1=0.8\,\Msun$ to $150\,\Msun$. Only below $z=2$ there is some variation, due to the increase in the production of solar-type stars, which are much more likely to be isolated.

It is notable that, relative to the AM model, the OB model leads to a significantly greater overestimation of the mass fraction in binaries than of the binary fraction. All else kept equal, this could be taken to suggest that, by treating all multiples as binaries, one would underestimate the number to mass ratio of binaries, leading to a lower merger rate density. Given the increase by a factor of $\sim1.2$ of the binary fraction, but $\sim10$ of the binary mass fraction, when going from the AM to the OB model, this underestimation could be by the order of a few. We thus suggest that, in the future, it will be interesting to compare the merger rates obtained in the OB case, and from the binaries in the AM case, in order to more accurately constraint the contribution from binary evolution to the total merger rate.

\subsection{Orbital parameters}
\label{sec4sub:orbital}

\begin{figure*}
    \centering
    \includegraphics[width=\textwidth]{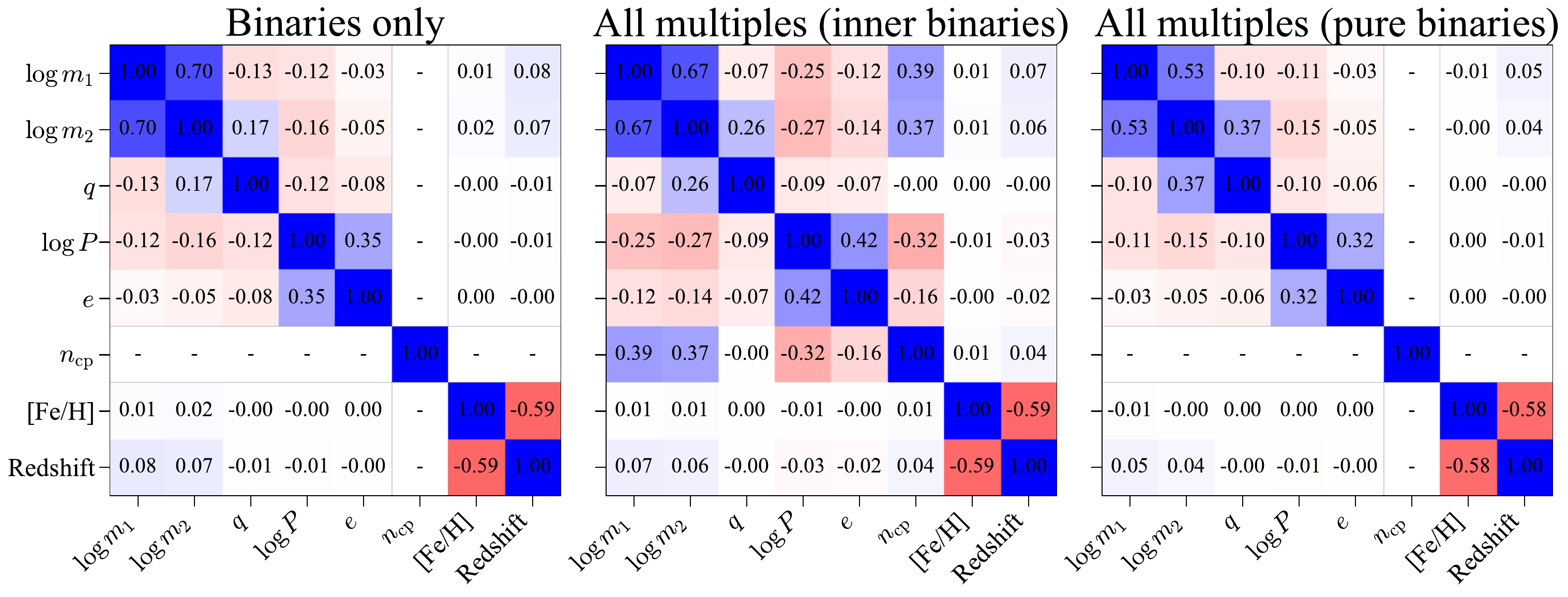}
    \caption{Kendall correlation matrices for $\log m_1$, $\log m_2$, orbital parameters, $\ncp$, redshift and metallicity, for binaries in the OB model (left column), inner binaries in the AM models (middle column) and physical binaries in the AM model (right column). Most (anti) correlations) are a direct consequence of our input distributions (Sec. \ref{sec:2distributions}). The stronger anticorrelation between orbital period/eccentricity and component masses for AM inner binaries is due to sampling bias of inner binaries. No significant correlations emerge between orbital parameters and redshift or metallicity.}
    \label{fig:correlations}
\end{figure*}

\begin{figure*}
    \centering
    \includegraphics[width=\textwidth]{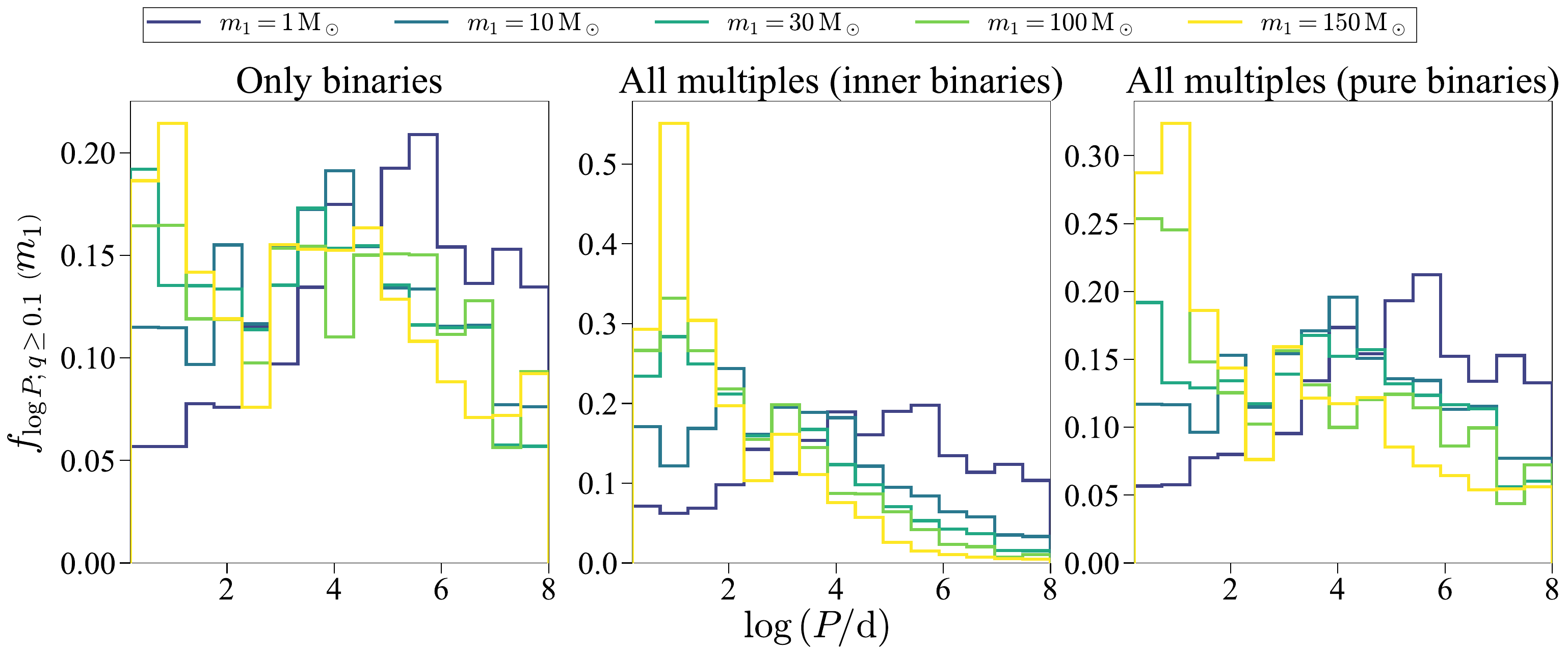}
    \caption{Orbital period distributions from samples of OB binaries, AM inner binaries and AM physical binaries. Each histogram is individually normalized so that its area equals unity, but they are otherwise equivalent to the companion frequency. OB binaries (left panel) are directly sampled from the companion frequency and retain its form. AM inner binaries (center panel) suffer a strong bias towards low periods by their very definition. AM physical binaries (right panel) are shifted towards low periods, relative to the original distribution, for high $m_1$ due to the rarity of binaries with very massive primaries (see Fig. \ref{sec2fig:allmult_frac}); this results in undersampling of the sub-dominant $\logP=3.4$ peak.}
    \label{fig:logp_distributions_mult}
\end{figure*}

Because $m_1$ is sampled from a redshift- and metallicity-dependent IMF, and all subsequent sampled parameters are correlated with $m_1$, a channel for a possible indirect dependence of those parameters on the environment is opened, even if their distributions are not explicitly dependent on it. We thus check for any such dependencies in the multiplicity fractions and orbital parameters.

We compute the correlation coefficient between redshift and metallicity and $m_1$, $m_2$, $\ncp$ and all orbital parameters; we opt for the Kendall coefficient because it does not make assumptions about the underlying shape of a possible correlation, but instead measures rank correlation, i.e., the similarity in the data when ranked by each variable in turn. We show the results in Fig. \ref{fig:correlations} for both AM and OB models, the latter both for the full set of inner binaries and the set of physical binaries; no expressive correlation between orbital parameters and redshift or metallicity emerges from linking the \citetalias{Jerabkova2018} IMF with the \citetalias{MoeDistefano2017} distributions.

The stronger anticorrelation between orbital period/eccentricity and component masses in the AM model inner binaries is a straightforward sampling bias. Inner binaries of high order multiples are by definition shifted towards shorter periods relative to the full companion population; because multiplicity increases with mass, this results in an anticorrelation between primary mass and the orbital period of the inner binary. As a direct consequence of the \citetalias{MoeDistefano2017} distributions (Fig. \ref{fig:md17_edistr}), this also leads to lower eccentricities. We show the resulting orbital period distributions (or the companion frequencies normalized to unity) for the three cases in Fig. \ref{fig:logp_distributions_mult}, where the OB case naturally retains the original shape (Fig. \ref{fig:md17_compfreq}), while the two others are shifted towards low periods. The strong relative shift at all masses for AM inner binaries is due to the overall sampling bias, while for the AM physical binaries we see the influence of the decreasing binary fraction with mass: for lower masses, where binaries are the most common multiple, the distribution stays close to the original shape (compare the left to the right columns); as $m_1$ increases, the distribution becomes increasingly shifted to short periods, relative to the original. This a reflection of the increasingly low number of physical binaries at very high masses, which fails to populate the sub-dominant $\logP=3.4$ peak.

\section{Conclusions}
\label{sec:5conclusions}

We have introduced \texttt{COMPAS}, a detailed initial sampling code for BPS, for either the Invariant initial parameter distributions often employed in BPS, or the more recent Varying IMF and correlated orbital parameter distributions. The pipeline generates a composite binary population, in which each simple population corresponds to a different metallicity-SFR-redshift bin. We are able to take into account higher-order multiples in the sampling, such that each systems is in the final sample defined by the environmental conditions of its formation (metallicity, SFR and redshift), the mass of its primary (the most massive component, $m_1$), the number of companions to the primary ($\ncp$), and the set of orbital parameters of each companions (orbital period, $P$; mass ratio, $q=\mcp/m_1$; and eccentricity, $e$). We compare the effects of the Varying IMF and correlated orbital parameters in compact object merger populations over time in an accompanying paper \citep{bossa2}. Here we have focused on the construction of \texttt{BOSSA}, and its consistency checks; and on discussion of nuances from star formation relevant to BPS.

With regard to mass and orbital parameters, we considered two sets of models: the Invariant model, where $m_1$ is sampled from a \citet{Kroupa2001} IMF; $P$ from the log-uniform distribution by \citet{OpikLaw}; $q$ from the uniform distribution by \citet{SanaMassRatio}; and $e=0$. In contrast, the Varying model feature $m_1$ is sampled from the \citet{Jerabkova2018} IMF, computed in the IGIMF framework and which becomes top-heavy for high SFRs, and top-heavy/bottom-light for low metallicities; and the orbital parameters are sampled from the correlated distributions presented by \citet{MoeDistefano2017}, which presumably reflect the pre-ZAMS evolution of multiple systems. We select the metallicity, SFR and redshift of the simple populations by a mass-weighted sampling of the metallicity-sensitive cSFH by \citet{Chruslinska2019}, and use the corrections to the SFR for the \citet{Jerabkova2018} IMF by \citet{Chruslinska2020}.

From the companion frequency as a function of $m_1$ and $\logP$ fitted by \citet{MoeDistefano2017} and their assumption of a Poissonian for the probability distribution of $\ncp$ for massive stars, we computed the multiple fractions as a function of $m_1$. We found  that, in order to reproduce the growth of the companion frequency extrapolated up to our maximum $m_1=150\,\Msun$, we need to allow up to $\ncp^\mathrm{max}=4$ companions. We thus obtain the fractions of isolated stars up to quintuple systems as a function of $m_1$ in the $\lrs{0.8\,\Msun,150\,\Msun}$ range. We define two models of multiplicity: in the All Multiples (AM) model, we sample all multiple systems up to quintuples from these fractions; in the Only Binaries (OB) model, we treat all multiples as binaries. The OB models yields the binary fraction as a monotonically increasing function of $m_1$ with mass-weighted average value of $0.74$ in the $\lrs{5\,\Msun,150\,\Msun}$ range, which is consistent with the constant $0.7$ fraction commonly assumed in BPS, based on \citet{SanaMassRatio}.

We also carefully consider the interpretation of the IMF with regard to mass sampling. We argue that, from an empirical perspective, the IMF as a distribution should be reproduced by the \textit{entire} set of component masses, regardless of their position within a system; and not necessarily by, e.g., the primary masses individually. We thus propose a pipeline in which all masses are initially sampled from the IMF, and in the following steps, companions are paired with primaries according to the chosen mass ratio distribution. We verify that these steps do reproduce the mass ratio distribution and the IMF simultaneously, when considering all sampled masses.

This also offers us a look into the question of how the IMF of different-order components vary relative to the full IMF. We model them by a two-part power law fit after finding that their variations can be generally divided into a lower-mass ($\leq10\,\Msun$) and a higher-mass ($>10\,\Msun$) range, and look at them in the model combinations of Invariant and Varying, and OB and AM. In the Varying model, the primary IMF is always flatter for lower masses and steeper for higher masses. The inverse behavior is obtained for the secondary IMF in the OB model, and by the tertiary and further IMFs, in the AM model. In the latter set, the secondary IMF is nearly the same as the full IMF for lower masses and steeper at higher masses. The relative behavior to the full IMF is maintained as it varies with metallicity and SFR/redshift. 

For the Invariant model combined with the OB model, both the primary and secondary IMFs stay close to the full IMF at lower masses, and become slightly flatter and steeper, respectively, at higher masses. The AM model shows greater variation: the primary IMF only deviates by becoming slightly steeper at low masses, and the secondary shows an attenuated version of the OB behavior; but all further companion IMFs are much flatter at all masses. We suggest that the shape of the component IMFs are set by two competing effects: the tendency of $\ncp$ to grow with $m_1$; and the tendency of companions with longer periods to be closer to $q=0.1$, driving their masses down. Because the Invariant model only features the first one, it shows a greater excess of massive companions; while this is attenuated by the second in the Varying model. In general, we find that companions will be less likely to be in the lower range due to correlations with primary mass and the increase of multiplicity with it.

We check for resulting variations of the multiplicity fractions with redshift in the Varying model, and find only an increase of about $0.1$ of the isolated fraction between $z=2$ and $z=0$ due to the decreasing SFR making the IMF less top-heavy, and thus amplifying the production of solar-type stars, which are much more likely to be isolated. This is present in both the OB or AM models. Because these are not compact object progenitors, this does not result in any significant variation for BCO population synthesis. We compute also the mass fractions, and find that, while number fractions decrease with $\ncp$, mass fractions increase. In the AM model, the binary mass fraction corresponds to less than $0.2$ of the total star-forming mass. In the OB model, in terms of CO progenitors, the correlation between $\ncp$ and $m_1$ leads to a binary mass fraction of nearly $0.9$ and a number fraction of $0.6$. This fact suggests that CO merger rates estimated by assuming a fixed number fraction of $0.7$ are overestimated due to being normalized by a lower star-forming mass.

Finally, we also check for any emerging correlations between orbital parameters and redshift or metallicity due to their coupling to a metallicity- and redshift-dependent IMF in the Varying model, but found no significant evidence for them. This is true for the cases of OB binaries, AM inner binaries (of all multiples) and AM physical binaries; significant differences between the three emerge due solely to sampling biases of AM inner binaries towards short periods (by definition); and to a scarcity of AM physical binaries with short periods, since such companion orbits are more common for massive primaries, which are preferably in triples or higher-order multiples.

These biases do indicate that the mass-dependence of multiplicity is a key factor in properly estimating binary merger populations, which should be characterized by very different masses and starting conditions then pairs of stairs within systems of arbitrary order; the latter might have significant implications for evolutionary channels of BCMs. While equating all multiples with binaries still offers an estimate of the \textit{total} merger rates from the isolated evolution of multiple systems, it is nevertheless important to take this mass-dependence into account in order to avoid errors when normalizing rates for the star-forming mass. Differentiating between different-order multiples, however, does offer us a chance of more accurately implementing binary evolution models just for physical binaries, and perhaps for inner binaries of higher-order multiples that evolve as if they were isolated, as done by \citet{Klencki2018}. In the case of BHNS and NSNS mergers, this can be expected to yield a good estimate for the total merger rate from isolated evolution, as NS progenitors should mostly originate in binaries. For BHBHs, however, the fraction of higher-order multiples participating in the merger rate might be significant, which is in line with the findings of \citet{Mapelli2021hierach,Vynatheya2022hierach}. As refined estimates of the merger rate contribution from higher-order multiples become available, separating true physical binaries for BPS is essential for properly combining these different contributions, and comparing the total merger rate to empirical constraints.

\section*{Acknowledgements}

We thank Martyna Chru\'sli\'nska for providing the grid of IGIMF corrections for the SFR. We would like to thank the anonymous referee for several suggestions that contributed to the accuracy and clarity of the paper.

 This paper made use of the BOSSA initial sampling code (version 1.0.0), which is avaiable at \url{https://github.com/lmdesa/BOSSA}. This research was funded by S\~ao Paulo Research Foundation (FAPESP) grant number 2020/08518-2. L.M.S. acknowledges funding from the National Council for Scientific and Technological Development (CNPq), grant number 140794/2021-2. L.S.R acknowledges funding from FAPESP, grant number 2023/08649-8. J.E.H. has been partially supported by the CNPq.

This work made use of the following software packages: \texttt{astropy} \citep{astropy:2013, astropy:2018, astropy:2022}, \texttt{Jupyter} \citep{2007CSE.....9c..21P, kluyver2016jupyter}, \texttt{matplotlib} \citep{Hunter:2007}, \texttt{numpy} \citep{numpy}, \texttt{pandas} \citep{mckinney-proc-scipy-2010}, \texttt{python} \citep{python}, \texttt{scipy} \citep{2020SciPy-NMeth}, \texttt{seaborn} \citep{Waskom2021} and and \texttt{PyTables} \citep{pytables}.

This research has made use of NASA's Astrophysics Data System.

Software citation information aggregated using \texttt{\href{https://www.tomwagg.com/software-citation-station/}{The Software Citation Station}} \citep{software-citation-station-paper, software-citation-station-zenodo}.
 
%%%%%%%%%%%%%%%%%%%%%%%%%%%%%%%%%%%%%%%%%%%%%%%%%%
\section*{Data Availability}

\texttt{BOSSA} and all associated post-processing code, including for generating plots included here, will be made available on GitHub upon publication. Data products underlying this article will be made available on Zenodo upon publication.

%%%%%%%%%%%%%%%%%%%% REFERENCES %%%%%%%%%%%%%%%%%%

% The best way to enter references is to use BibTeX:

\bibliographystyle{mnras}
\bibliography{bibliography} % if your bibtex file is called example.bib

% Alternatively you could enter them by hand, like this:
% This method is tedious and prone to error if you have lots of references
%\begin{thebibliography}{99}
%\bibitem[\protect\citeauthoryear{Author}{2012}]{Author2012}
%Author A.~N., 2013, Journal of Improbable Astronomy, 1, 1
%\bibitem[\protect\citeauthoryear{Others}{2013}]{Others2013}
%Others S., 2012, Journal of Interesting Stuff, 17, 198
%\end{thebibliography}

%%%%%%%%%%%%%%%%%%%%%%%%%%%%%%%%%%%%%%%%%%%%%%%%%%

%%%%%%%%%%%%%%%%% APPENDICES %%%%%%%%%%%%%%%%%%%%%

%%%%%%%%%%%%%%%%%%%%%%%%%%%%%%%%%%%%%%%%%%%%%%%%%%

% Don't change these lines
\bsp	% typesetting comment
\label{lastpage}
\end{document}